\documentclass[12pt,preprint]{aastex}

\newcommand{\etal}{{et al.~}}

\slugcomment{Submitted to the Astrophysical Journal}
\shorttitle{Magnetic Field in OMC 2}
\shortauthors{Poidevin, F.; Bastien, P.; Matthews, B. C.}

\begin{document}  

\title{MAGNETIC FIELDS STRUCTURES AND TURBULENT COMPONENTS IN THE STAR FORMING MOLECULAR CLOUDS OMC-2 AND OMC-3}
\shorttitle{}
\shortauthors{}


\author{Fr\'ed\'erick POIDEVIN}
\affil{Universidade de S\~{a}o Paulo, Instituto de Astronomia, Geof\'isica e C\^ien\c{c}as Atmosf\'ericas, 
Rua do Mat\~{a}o 1226, Butant\~{a}, S\~{a}o Paulo, SP 05508-900, Brazil}
\email{Poidevin@astro.iag.usp.br}

\author{Pierre BASTIEN}
\affil{D\'epartement de Physique and Observatoire du Mont-M\'egantic, Universit\'e de Montr\'eal, 
C.P. 6128, Succ. Centre-ville, Montr\'eal, Qu\'ebec H3C 3J7, Canada}
\email{Bastien@astro.umontreal.ca}

\author{Brenda C. MATTHEWS}
\affil{Herzberg Institute of Astrophysics, National Research Council of Canada, Victoria, BC,
Canada}
\email{brenda.matthews@nrc-cnrc.gc.ca}

\begin{abstract}

The SCUBA polarized $850$ $\mu$m thermal emission data of the region OMC-2 in Orion A are added to 
and homogeneously reduced with data already available in the region OMC-3. 
The data set shows that OMC-2 is a region generally less polarized than OMC-3.
Where coincident, most of the $850$ $\mu$m polarization pattern is similar to that measured in $350$ $\mu$m polarization data. 
Only $850$ $\mu$m polarimetry data have been obtained in and around MMS7, FIR1 $\&$ FIR2, and in the region south of FIR6. 
A realignment of the polarization vectors with the filament can be seen near FIR1 in the region south of OMC-3. 
An analysis shows that the energy injected by CO outflows and H$_{2}$ jets associated to 
OMC-2 and OMC-3 does not appear to alter the polarization patterns at a scale of the $14''$ resolution beam. 
A second order structure function analysis of the polarization position angles 
shows that OMC-2 is a more turbulent region than OMC-3. 
OMC-3 appears to be a clear case of a magnetically dominated region with respect to the 
turbulence. However for OMC-2 it is not clear that this is the case. 
A more in-depth analysis of five regions displayed along OMC-2/3 indicates a decrease of the 
mean polarization degree and an increase of the turbulent angular dispersion from north to south. 
A statistical analysis suggests the presence of two depolarization regimes in our maps.  
One regime including the effects of the cores, the other one excluding it. 
\end{abstract}


\newpage

\section{INTRODUCTION}
 
  It is generally believed that magnetic fields play an important role in star
formation but the evolution of their role from the scale of molecular clouds
to that of young stellar objects is still not well understood. One of
the best methods for their study is polarimetry of the continuum radiation emitted by aligned dust grains
at far-infrared (FIR) to submillimeter (submm) wavelengths \citep[e.g.][]{dot00,hil99,mat09}.
Due to fast rotating mechanisms, elongated grains pervading the dense ISM should mostly have their
longer axis preferentially oriented perpendicular to the local magnetic
field. Recent advances on grain alignment theory can be found in reviews by
\citet{laz03} and \citet{laz07}.
 
  The Orion ``Integral-Shaped filament'' (ISF) was mapped
at $850$ $\mu$m by \citet{joh99}. It contains the well-studied OMC-1 region 
behind the Orion nebula at $\approx 414$ pc \citep[see][]{men07}, and the two filamentary structures
OMC-2 and OMC-3 located $\approx 15'$ and $\approx 25'$ north of OMC-1, respectively.
Submm continuum and polarimetry observations in OMC-1 were intensively analysed 
and discussed by \citet{kee82}, \citet{hil84}, \citet{ait97}, \citet{lis98},
\citet{rao98}, \citet{sch98}, \citet{val99}, \citet{cop00} and \citet{val07}.
The counterpart of these studies in OMC-2 and more particularly in OMC-3
were discussed by \citet{chi97} with $1.3$ mm dust emission observations in
both regions; by \citet{mw00} and \citet{mwf01} with $850$ $\mu$m polarimetry
in OMC-3; by \citet{hou04} with polarimetry at $350$ $\mu$m in both
filaments; and more recently by \citet{mat05} with $1.3$ mm high resolution observations
in the MMS6 protostellar core in OMC-3.
 
The 850 $\mu$m polarization pattern of OMC-3 measured with SCUBA
and studied by \citet{mwf01} shows that a helical magnetic field \citep[see][]{fp00d}
could thread the filament. The change in orientation between the polarization data and the
filamentary axis to the south of OMC-3 could be accounted for by either a bend of 
the filament or the presence of a second filament, oriented roughly orthogonal to the primary ISF. 
If the latter interpretation is correct, then polarization vectors realigning with the
filamentary axis are expected south of OMC-3. If the filament
continues to bend, then any orientation of vectors to the filament axis is possible.

Our initial motivation for this project was to measure
the 850 $\mu$m polarization pattern of OMC-2 with the aim
to better understand the structure of the magnetic field south
of OMC-3. Then, with the availability of 350 $\mu$m polarization data
by \citet{hou04} and estimates of the orientation
of the magnetic field relative to the Line-of-Sight (LOS) at several places along
the filament, a global comparison with the 850 $\mu$m 
polarization vectors observed in both regions can be made. 
Abundances of several molecular species and clumping were also observed
in these regions \citep[e.g.][]{bat83,cas95,chi97} suggesting an evolutionary effect 
from north to south along OMC-3/2. 
In contrast, \citet{tak08} show that some intermediate-mass (IM) star-forming objects in OMC-3 
can be at a more evolved stage than some less active IM star-forming objects in OMC-2. 
Observations of H$_{2}$ jets \citep[e.g.][]{sta02,par06} 
and of molecular outflows \citep[e.g.][]{wil03}
allowed the detection of some progenitors along the chain of
active cores embedded in the filament with relatively good certainty. 
Thus we can compare their respective orientations on the plane of the sky (POS) 
with the direction of polarization vectors in order to statistically study the 
impact of outflow and jet activity on the region. 
The distribution of the offsets in position angles (P.A.s) obtained in this way can  
help to test detectability of turbulence in the framework of our observations.
In addition, a coherent structure function analysis method was recently put forth 
by \citet{hil09}. Combined with other analysis tools, this promising approach should help 
to make fruitful comparisons with available and future models. As a consequence we can use 
this method along the OMC-2 and OMC-3 filamentary molecular clouds and compare it with 
steady state and turbulent magnetic field models. 
All in all, these approaches should help to better understand the impact of magnetic fields
with respect to turbulence and gravity in star-forming processes and cloud evolution. 

In this work we present 850 $\mu$m SCUBA polarization data of OMC-2.
Observations and data reduction techniques are presented and discussed
in $\S$ \ref{OBS}. Results are shown and analyzed in $\S$ \ref{RES}.
A discussion and comparisons with models follow in $\S$ \ref{DISCU}.

\section{OBSERVATIONS AND DATA REDUCTION} \label{OBS}

The observations were carried out on the summit of Mauna Kea in Hawaii, from 2004 
September 14 to 20 and on 2005 January 16  
with the SCUPOL polarimeter installed in front of the Submillimeter Common User Bolometric Array (SCUBA) 
detector (\citet{hol98}) mounted on the James Clerk Maxwell Telescope (JCMT). 
The nights were relatively stable with $\tau$ (225 GHz) ranging from 0.04 to 0.07 during 
the periods of observations. The polarimeter and general reduction techniques 
are described by \citet{gre03}.
 
The SURF \citep[SCUBA User Reduction Facility; see][]{jen98} and KAPPA 
reduction packages were used for flatfielding, extinction correction, instrumental polarization 
removal, sky noise removal, despiking, removal of bad pixels and rebinning of the images.  
The Starlink software packages POLPACK and CURSA were also used for combining, binning and filtering 
the Stokes parameters of the images to finally extract the polarized source signal.

After extinction correction, noisy bolometers were identified and removed from the data sets.
At 850 $\mu$m, effects due to sky variations  
were subtracted by using bolometers devoid of significant target flux. 
We used between one and four bolometers to determine sky variability, using the existing 
850 $\mu$m map of total intensity of the ISF produced by \citet{joh99} to help select 
empty bolometers. Bolometers which were significantly negative were avoided   
since they can greatly affect the interpretation of the
final map by possibly injecting polarized emission in the bolometers probing the 
source \citep[See Appendix B of][for more details]{mwf01}.  
The mean flux removed by sky subtraction was added back into 
the maps since the flux from sky emission was not necessarily always close enough to zero.

During observations, the atmospheric variability was cancelled by chopping. The pointing center,    
the chop position angle, the distance to the center of the chop position and the number of 
observations for each of the seven different SCUBA fields required to map the OMC-2 filament
and the southernmost portion of OMC-3 are shown in Table \ref{JIG}.  

  After removal of the instrumental polarization in each bolometer, the Stokes parameters 
$Q$, $U$ and $I$ for each set of data were combined to produce a final cube of data where $Q_{tot}$ and
$U_{tot}$ are the two components of linearly polarized light and $I_{tot}$ is
the total intensity associated with each $6.18''$ pixel of the map. The polarization 
percentage, $p$, and the polarization position angle, $\theta$, are respectively defined according 
to the usual relations

\begin{equation} \label{p}       
p= \frac{\sqrt{(Q^{2}+U^{2})}}{I}, 
\end{equation}
\begin{equation}  \label{theta}       
\theta=\frac{1}{2} \arctan \left(\frac{U}{Q}\right).
\end{equation}

The uncertainties on $p$ and $\theta$ are respectively given by
\begin{equation}  \label{sigp}       
\sigma_{p}=\frac{\sqrt{dQ^{2}Q^{2}+dU^{2}U^{2}}}{p}, 
\end{equation}
\begin{equation}  \label{sigtheta}       
\sigma_{\theta} = 28.6^{\circ}\left(\frac{p}{\sigma_{p}}\right)^{-1},
\end{equation}
where $\frac{p}{\sigma_{p}}$ is the signal-to-noise ratio (S/N) in $p$.
The polarization defined in equation (\ref{p}) is overestimated and 
needs to be debiased using the relation

\begin{equation}  \label{pdebiased}       
p_{\rm db} = \sqrt{p^{2}-\sigma_{p}^{2}}.
\end{equation}
In what follows we refer to $p_{\rm db}$ and we call it $p$.

OMC-2 data were combined with SCUBA data of OMC-3 \citep[see][]{mw00,mwf01} 
retrieved from the JCMT data archive. Since the data reduction does not provide 
an absolute sensitivity to the polarization intensity appropriate values of the Flux 
Calibration Factors (FCFs) were defined by comparing the intensity flux levels of the 
OMC-2 and OMC-3 intensity maps with the ISF continuum map of \cite{joh99} and the OMC-2/3
map of \citet{dif08}.   
This ensures a reasonable similarity between the OMC-3 and OMC-2 data sets.
The values of the FCFs employed were 484 Jy/beam/V in OMC-3 and 690 Jy/beam/V in OMC-2.

Overestimation of the S/N ratio due to oversampling in the pixel scale was minimized during the reduction
process by sampling to a larger grid than that used by \citet{mwf01}. This means that our reduction is
not exactly comparable to the earlier publication. Similarly, our reduction is not identical to that of
\citet{mat09} (the SCUPOL Legacy Catalogue). Despite these differences in sampling, the results are 
generally consistent with one another. 

The data for the eleven fields were reduced 
homogeneously and combined to produce a preliminary 850 $\mu$m polarization map covering both regions.
In order to study the variations in the polarization patterns, this preliminary map 
was binned by factors of 1, 2, 3, 4 and 5 pixels, and each one of these maps was filtered
with the following criteria $I_{tot} > 0$ and either $\sigma_{p} < 0.75 \%$ or $\sigma_{p} < 1.0 \%$.
The ten maps produced by these various combinations were compared with each other
and we decided that the map binned to $\approx 18.6''$ (or $\approx 1.3 \times 14 ''$ beam resolution)  
improves $p/\sigma_{p}$ and hence $\sigma_{\theta}$ sufficiently. 
This rebinning improves S/N by a factor of 3 over the unbinned $6.18''$ sampled data.
The filtering criterion $\sigma_{p} < 0.75 \%$ was then chosen because it avoids
polarization artefacts in low emissivity regions due to binning effects. 

\section{RESULTS} \label{RES}  

\begin{figure}
\epsscale{0.6}
\plotone{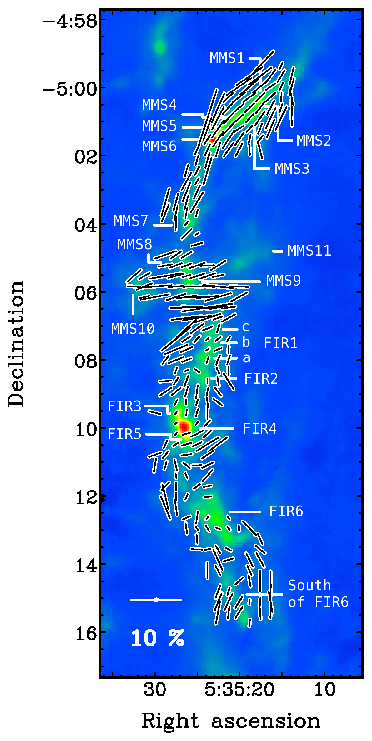}
\caption{850 $\mu \rm m$ map and polarimetry of the OMC-3 and OMC-2 regions 
obtained with SCUBA at the JCMT. The vector length denotes polarization percentage and 
the position angle denotes the E-vector orientation. The vectors have a polarization percentage and 
uncertainty such that $p/\sigma_{p} > 3$. 
Coordinates are J2000.0.  
\label{polmap}}
\end{figure}

\begin{figure}
\epsscale{0.5}
\plotone{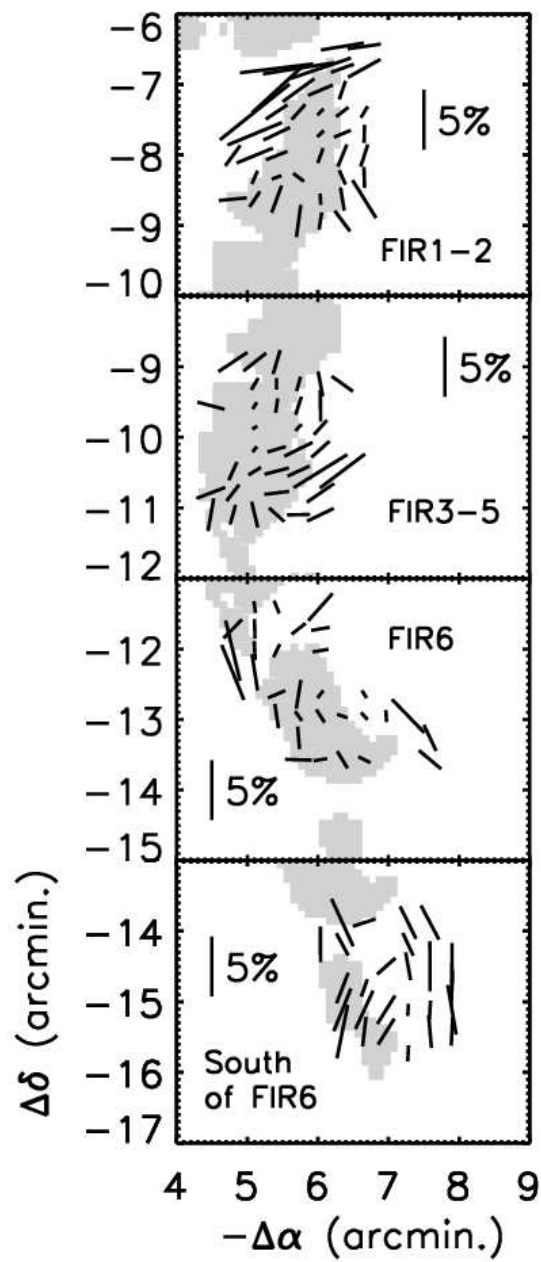}
\caption{
Zoom on the four regions located along OMC-2 and displayed in Table \ref{MEANDISP}. 
All the data shown here have a signal-to-noise ratio such that $p/\sigma_{p}>3$. 
A background intensity map ($I(850 \mu m) \ge 0.5$ Jy/pixel, where pixels are 
$3 \arcsec \times 3 \arcsec$ in size) is displayed and shows the location of the filament.
The reference position is R.A.=5$^{\rm h}$35$^{\rm m}$23.5$^{\rm s}$, decl.=-5$^{\circ}$ 01$^{\rm mn}$ 32.2$^{\rm s}$ (J2000.0).   
See Figure \ref{polmap} for comparisons.
\label{mapsubregions}}
\end{figure}

The final 850 $\mu$m polarization map of OMC-2/OMC-3 is shown in Figure \ref{polmap}.
We present detailed maps of four subregions of OMC-2 in Figure \ref{mapsubregions}.
We follow the same notation as \citet{chi97} in their Figure 
to identify the cold condensations encountered along the filaments;
they are also identified in our Figure \ref{polmap}. 
The vectors shown in the figures have $p/\sigma_{p} > 3$.

\subsection{Degree of Polarization and Distribution of Polarization Position Angles} \label{deg}

The polarization data along OMC-2 are reported in Table \ref{DATA}.
Position angles (P.A.) are counted positively east of north.
The results of a basic statistical analysis on several subsets of the sample are shown in Table  
\ref{MEANDISP} where mean position angles, $<$P.A.$_{\rm fila}>$, of the corresponding sections of 
the filaments are displayed. These mean orientations within the filament were estimated 
by eye with the help of contour maps.    

\begin{figure}
\epsscale{0.8}
\plotone{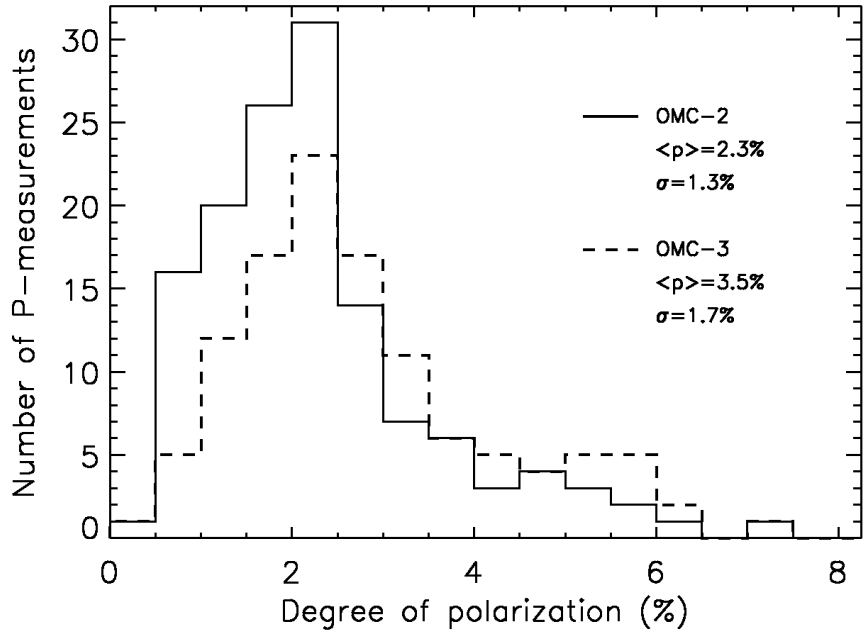}
\caption{Histogram of 850 $\mu$m degree of polarization through OMC-3 and OMC-2
based on data shown in Figure \ref{polmap}. 
\label{phistot}}
\end{figure}

The histogram of the OMC-2 and OMC-3 polarization percentages data sets are shown in Figure \ref{phistot}. 
The mean polarization of the combination of the two data sets is $<p> = 2.8 \%$ with a dispersion of $1.6 \%$. 
Within the uncertainties, no polarization vector greater than $12 \%$ can be seen in the distribution. 
Vectors within OMC-3 show a mean polarization percentage of $<p> = 3.5 \%$, 
slightly higher than the mean polarization percentage of $<p> = 2.3 \%$ found within 
OMC-2.

\begin{figure}
\epsscale{0.8}
\plotone{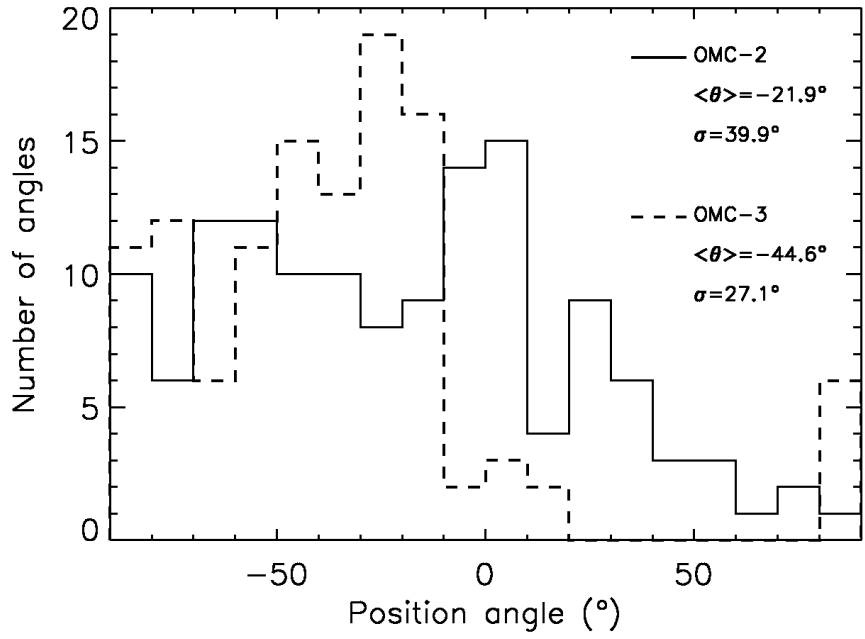}
\caption{Histogram of 850 $\mu$m polarization position angles through OMC-3 and OMC-2
based on data shown in Figure \ref{polmap}. 
\label{anghistot}}
\end{figure}

The histogram of the P.A. distributions in OMC-2 and OMC-3 are shown in Figure \ref{anghistot}.
The combination of the distributions peaks around a P.A. of $\approx -35^{\circ}$ with a dispersion $\sigma_{p} 
\approx 36^{\circ}$. The OMC-3 data peaks around a P.A.of $\approx -45^{\circ}$ while 
the OMC-2 data peaks around a P.A. of $\approx -22^{\circ}$. 
The distribution of vectors within OMC-3 has a smaller dispersion, $\sigma_{p} \approx 27^{\circ}$, than the one 
within OMC-2 where $\sigma_{p} \approx 40^{\circ}$, in agreement with the more ordered polarization pattern observed 
in this region.

\begin{figure}
\epsscale{1.0}
\plotone{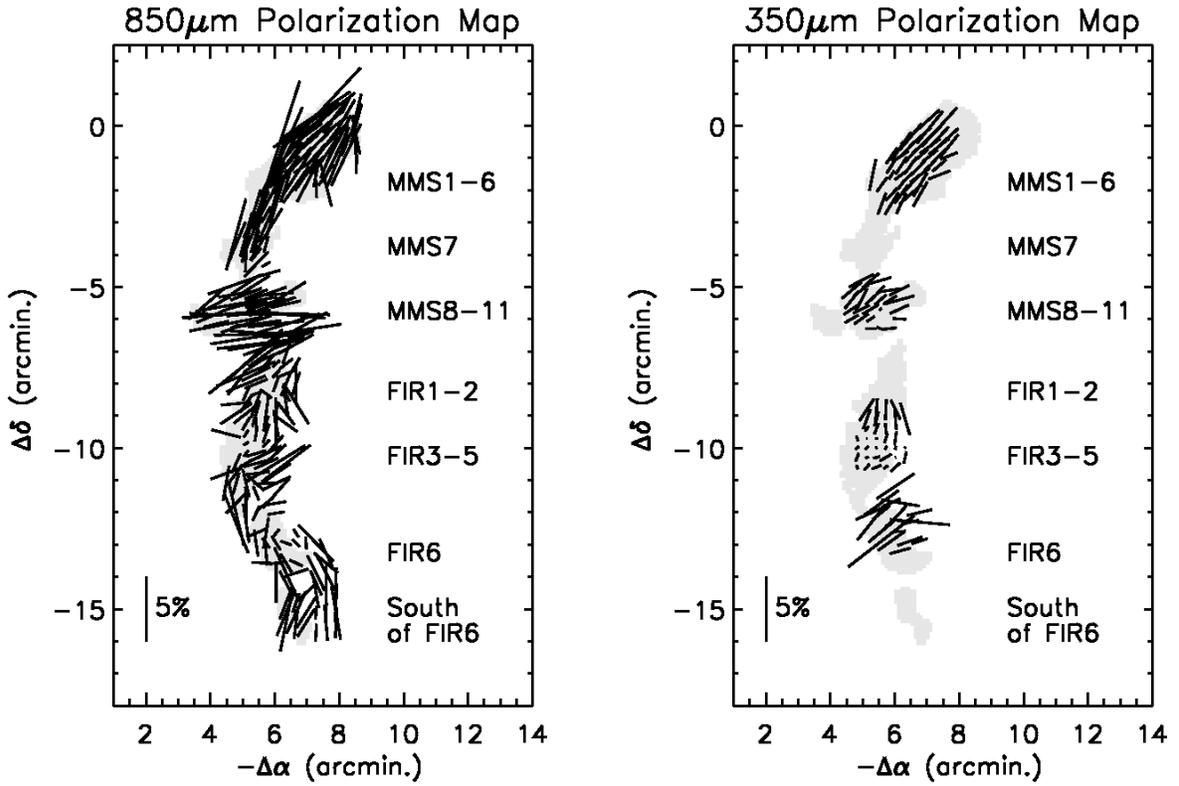}
\caption{Left: 850 $\mu$m polarization map from this work.
Right: 350 $\mu$m data from Houde et al. (2004). 
All the data shown here have a signal-to-noise ratio
such that $p/\sigma_{p}>3$. Each set of data is drawn with the same 
scale for comparison. A background intensity map ($I(850 \mu m) \ge 0.5$ Jy/pixel, where pixels are 
$3 \arcsec \times 3 \arcsec$ in size) is displayed and shows the location of the filament.
The reference position is R.A.=5$^{\rm h}$35$^{\rm m}$48$^{\rm s}$, decl.=-5$^{\circ}$ 00$^{\rm mn}$ 00$^{\rm s}$ (J2000.0).   
\label{mapcomp}}
\end{figure}

Figure \ref{mapcomp} compares the 850 $\mu$m data with the 350 $\mu$m data of \citet{hou04}.
Only $850$ $\mu$m data have been obtained in and around MMS7, FIR1 $\&$ FIR2 and in the region South of FIR6,
while data at both wavelengths are available in other regions.
The similarity between the polarization patterns can clearly be seen in most of the common areas where vectors 
at both wavelengths are detected. The same scale is used to plot the vectors, we also see that the polarization 
percentage is generally higher at $850$ $\mu$m than at $350$ $\mu$m.
We now discuss more specifically OMC-2 regions as designated on Figure \ref{polmap}. 

\subsubsection{FIR1 $\&$ 2}

\citet{hou04} only show polarization data southward of FIR2. 
The data of Figures \ref{polmap} and \ref{mapsubregions} contain 
the first submm data polarimetry in this region.
Polarization is detected toward FIR1 and FIR2 and along both sides of the filament. 
Depolarization is clearly visible from low to high column density in the direction of the densest regions.
The filament is oriented at an angle of $\approx -17^{\circ}$.  
The mean P.A. of vectors covering FIR1 and FIR2 is $-41.5^{\circ}$ with a large dispersion of $35.6^{\circ}$.
This mean P.A. is offset by $\approx 25^{\circ}$ with respect to the mean orientation 
of the filament in this region. Our vectors located in the region south of FIR2 show a polarization pattern 
consistent with the orientation of the 350 $\mu$m vectors shown in Figure 2 of \citet{hou04}. 

More generally, we point out that vectors located around and to the north of FIR1, although shifted
by about $30^{\circ}$ with the mean orientation of the filament in this region, 
have approximately the same orientation as vectors in the southern part of OMC-3.
We also note that the high angular dispersion $s(\theta) = 35.6^{\circ}$ of the data set covering FIR1 
and FIR2 implies a fast variation in position angles in a region having a spatial scale comparable 
in size with the region formed by MMS1 to MMS6 where the dispersion is approximately half of this value. 
Finally, the mean degree of polarization covering FIR1 and FIR2   
is $2.5 \%$ with a dispersion of $1.6 \%$.

\subsubsection{FIR3, 4, $\&$ 5}

The polarization pattern is consistent with that found by \citet{hou04} at 350 $\mu$m.
The vectors north of FIR3 have P.A. $\approx 0^{\circ}$ while a 
clear rotation of $\approx 90^{\circ}$ can be seen when moving south of FIR5. 
This trend gives a mean P.A. of $-47^{\circ}$ over the region and a dispersion 
of $\approx 35^{\circ}$. The filament is oriented at an angle of $\approx -10^{\circ}$ on 
the POS. The mean level of polarization  $<p> = 2.0 \%$ 
is the second lowest of the several subsets of data shown in Table \ref{MEANDISP} 
and the dispersion of the subset is $1.0 \%$.      

\subsubsection{FIR6}

Few detections are found in the densest emitting regions and the mean P.A. of
$26.2^{\circ}$ may indicate a mean alignment of the vectors with the filament 
which is oriented at $\approx 30^{\circ}$. 
Figure \ref{mapcomp} shows that the 850 $ \mu$m vectors are 
however generally inconsistent with the polarization 
pattern at 350 $\mu$m \citep[][]{hou04}. Moreover, degrees of 
polarization measured at 850 $ \mu$m appear to be about two to three times smaller
than those measured at 350 $\mu$m. As mentioned above, our mean P.A. in the region is  
$26.2^{\circ}$ with a dispersion of $44.5^{\circ}$ while, based on average Stokes parameters, 
\citet{hou04} found a mean P.A. of $\approx -65^{\circ} \pm 6.0^{\circ}$. 
The reason for this difference is not obvious.
The lower polarization percentage values are more prone to P.A. errors, for instance due to chopping onto polarized emission. 
There is no easy way to check this though.
Finally, at 850 $\mu$m, the mean degree of polarization of $1.9 \%$ is the lowest of all the subsets 
of data shown in Table \ref{MEANDISP} with a dispersion of $1.1 \%$. 

\subsubsection{South of FIR6} \label{southfir6}

These measurements are the first to probe aligned grains with submm polarimetry  
in this region. The vectors are relatively well aligned around a mean P.A. of $-5.7^{\circ}$
with a dispersion of $24.1^{\circ}$ and are offset by $\approx 27^{\circ}$ from the ridge orientation. 
The mean degree of polarization is $2.8 \%$ with a dispersion of $1.0 \%$.

\subsubsection{OMC-3}  \label{OMC-3}

The 850 $\mu$m data shown by \citet{mwf01} were already compared with 350 $\mu$m data by \citet{hou04}; therefore,
we will not discuss this region in detail here. 
Table \ref{MEANDISP} contains the mean and dispersion of polarization percentages of several regions of OMC-3.
We point out however that while MMS10 is located at approximately the same
declination as MMS8 and 9 but at a position $\approx 1'30''$ to the east, another dense condensation 
with no apparent star-forming core can be seen to the west of MMS8 at approximately the same distance. 
In Figure \ref{polmap} we have identified this faint condensation by ``MMS11''.  

With the data in OMC-2, we now see that a realignment of the polarization 
pattern with the filamentary structure is effective just to the south of OMC-3, near FIR1.
Data are not available in MMS11 but this realignment suggests that there 
is effectively a second filament crossing the region located from MMS10 to MMS11. 
We point out however that such a realignment does not reject
the possibility of a double bend of the filament with one bend located to the north
of the region MMS8 or around the region MMS10, and with 
the other one located to the south of the region MMS11 (see \citet{mwf01} and \citet{fp00d} 
for more details).  

\subsection{Polarization Hole} \label{hole}

\begin{figure}
\epsscale{1.0}
\plotone{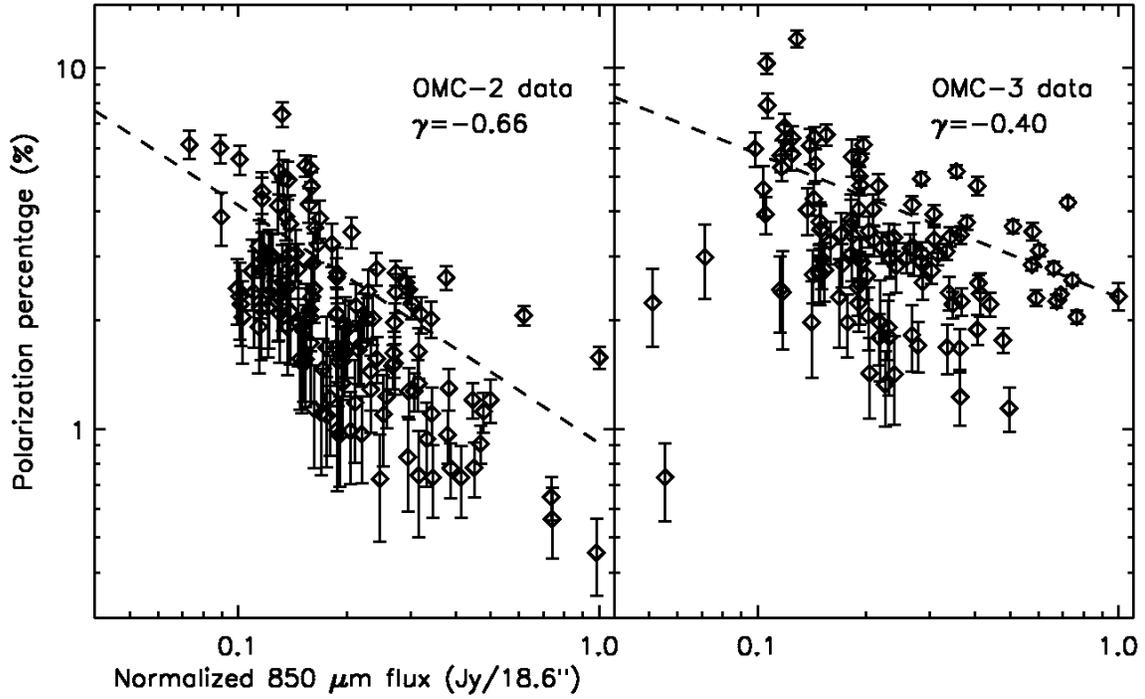}
\caption{Percentage of polarization versus 850 $\mu$m normalized continuum flux of the 
polarization data set in OMC-3 and OMC-2 shown in Figure \ref{polmap}. 
All the data shown here have a polarization percentage and uncertainty such that $p/\sigma_{p} > 3$ (see Figure \ref{polmap}). 
The dashed lines are $\chi^{2}$ power-law fits of the form $p = AI^{\gamma}$.
\label{pvsi}}
\end{figure}

In studies of polarized radiation emitted by aligned dust grains, it is usual to show the 
distribution of polarization as a function of intensity. Many of these distributions 
show a depolarization effect where $p$ decreases when $I$ increases.
Such an effect was shown by \citet{mwf01} in OMC-3, and Figure \ref{pvsi} shows that it can also 
be seen in OMC-2. The dashed lines are $\chi^{2}$ power-law fits 
of the form $p = AI^{\gamma}$ to the OMC-2 (left) and OMC-3 (right) data sets. 

We have calculated the power index $\gamma$ for several subsets of the data shown in Figure \ref{polmap}. 
The division of the data into subsets 
was partially arbitrary but such that the number of pixels is still statistically significant for each subset. 
In each data set, the flux was normalized by its maximum flux value in the data set.
Values of the power index are displayed in column (3) of Table \ref{subregions}. 
The number of pixels used in the fit for each data set are given in column (2) of the table.
One can see strong variations of the power index from one region to the other. 
The region MMS1 to MMS7 is distinguished from the region MMS8 to MMS10 based on the fact that 
the polarization patterns have different mean orientations. 
Given the strong flux emission along certain lines of sight, the region of FIR 1 and FIR 2 was distinguished 
from the region of FIR 3 to FIR 5. The region of FIR 6 was combined with the southernmost part of the map; 
these two areas contain a lot of pixels where no polarization was detected.
This may introduce a bias in the estimation of the power index and one should be cautious in the interpretation of the results
associated with these two areas.

\subsection{CO Outflows and H$_2$ jets} \label{jetsout}

\citet{wil03} imaged CO outflows at $10''$ resolution near
several protostellar sources in OMC-2 and OMC-3. The outflow properties 
including their lengths are summarized in their Table 1. We have directly 
used Figures 3 to 6 of their work to estimate the central coordinates and 
position angles of the molecular outflows spread along the filaments.
Unbiased H$_{2}$ surveys for protostellar jets in Orion A were also conducted 
by \citet{sta98,sta02}. The positions, P.A.s, and lengths of the 
H$_{2}$ features are listed in the Table 3 of \citet{sta02}.
Since it was shown by \citet{yu00} that some of the jets can not be associated with  
OMC-3 cores, only those associated with CS cores in OMC-3 \citep[see][]{tat1993,aso2000} 
are considered in the following analysis.

\begin{figure}
\epsscale{1.5}
\plotone{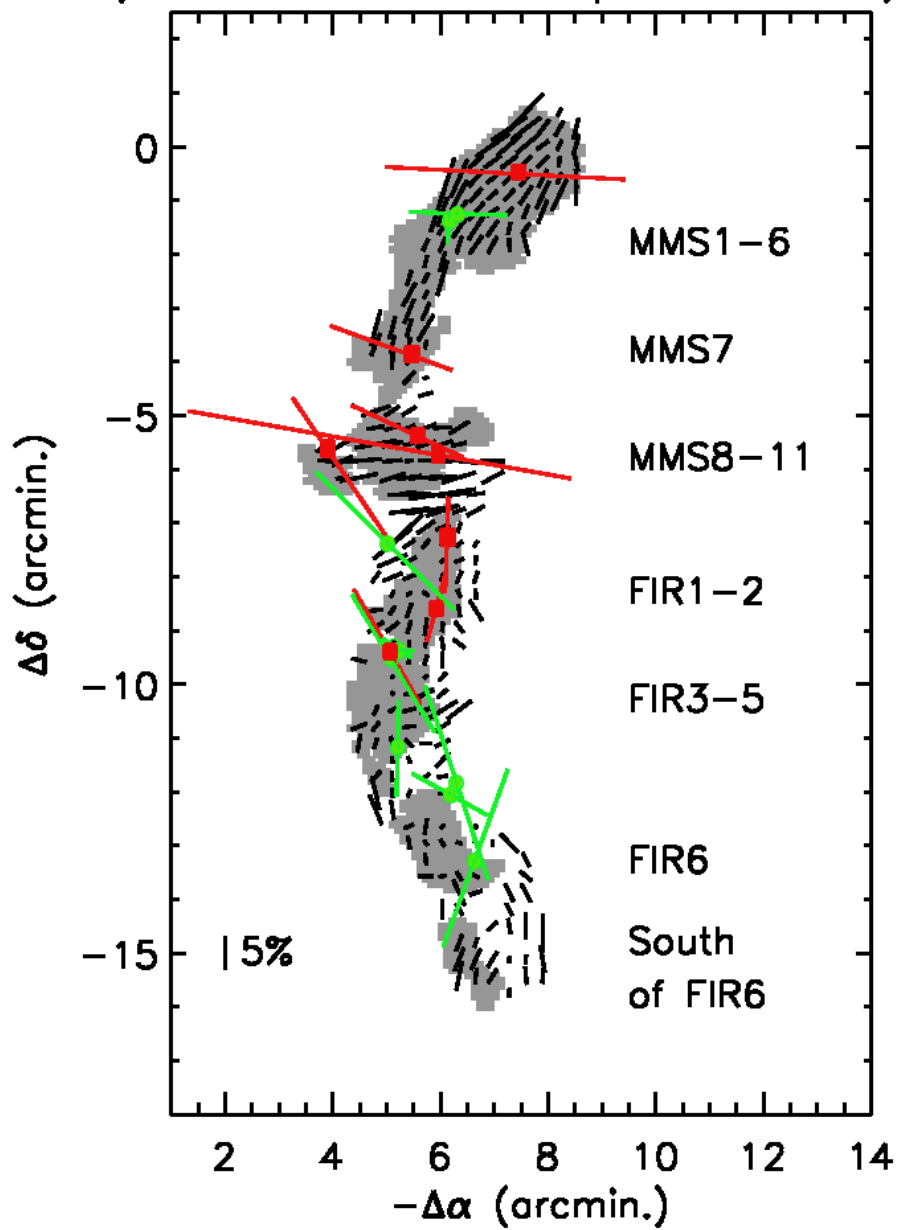}
\caption{Distribution of H$_{2}$ jets (green) and CO outflows
(red) superimposed on 850 $\mu$m polarization data (thin vectors). 
See section \ref{jetsout} for explanations and Figure \ref{polmap} for comparison with polarization vectors only.
Some of the H$_{2}$ jets over-plot CO outflows which make them difficult to see in the Figure. 
Flow designations and coordinates are given with other information in Table \ref{JETS}. 
A background intensity map ($I(850$ $ \mu \rm m) \ge 0.5$ Jy/pixel, where pixels are 
$3 \arcsec \times 3 \arcsec$ in size) is displayed and shows the location of the filament.
The reference position is R.A.=5$^{\rm h}$35$^{\rm m}$48$^{\rm s}$, decl.=-5$^{\circ}$ 00$^{\rm mn}$ 00$^{\rm s}$ (J2000.0).    
\label{jo}}
\end{figure}

All existing data associated with driving sources in OMC-2 and OMC-3 were used with the 850 $\mu$m data
to produce a map where the polarization pattern is superimposed on jets and outflow P.A.s. 
Figure \ref{jo} shows such a map where CO outflows and H$_{2}$ jet lengths are appropriately scaled.
CO outflows originating from MMS2/MMS3, MMS5, MMS7, MMS8, MMS9, MMS10, 
FIR1bc, FIR2 and FIR3 are shown as red lines and marked with a red full square symbol. 
This symbol shows the location of the \citet{chi97} dust condensations, but because 
many possible progenitors can be seen through FIR1bc and FIR2, 
they only indicate the geometrical centers of these flows. 
H$_{2}$ jets 4, 5, 14, 17, 18, 19, 21, 23, 24 and 25  \citep[][]{sta02}
associated with CS cores are shown with green lines and marked with a green full circle symbol, 
following the coordinates given in columns 2 $\&$ 3 of Table 3 of \citet{sta02}. 
All of the H$_{2}$ jets included here are members of the ``certain'' group
defined by \citet{sta02} where flows are identified on morphological 
grounds. The coordinates are thus representative coordinates, 
meaning that if there is a candidate driving source, its position 
is given; alternatively, if the location of a possible driving source 
is suggested on morphological grounds like for example, the geometric center 
of an apparently bipolar H$_{2}$ configuration, this position is given.     

On this basis a statistical study of the angular offsets on the POS between jets or 
outflows P.A.s and the P.A.s of the polarization vectors was carried out. 
The closest vectors on either side along the projected lengths were selected
with the distance between polarization vectors and flows being never higher 
than 18.6$\arcsec$; more typically, the distance is a few $\arcsec$. Each selected vector was used 
no more than one time. Each resulting mean angular offset is thus 
the difference between the P.A. of the outflow/jets and the mean of 
subsets of two to nine polarization P.A.s. Finally, we note that no 
jets are seen in the region south of FIR6  \citep[][]{sta02} and that the CO observations 
of \citet{wil03} did not cover this region. 
A summary of our results is given in Table \ref{JETS} where designations of the flows, 
the coordinates of the central source or of the central position, 
estimates of their P.A.s, estimates of the mean P.A.s
of subsets of submm vectors in their vicinity and offsets 
between these P.A. estimates are displayed. 

\begin{figure}
\epsscale{0.8}
\plotone{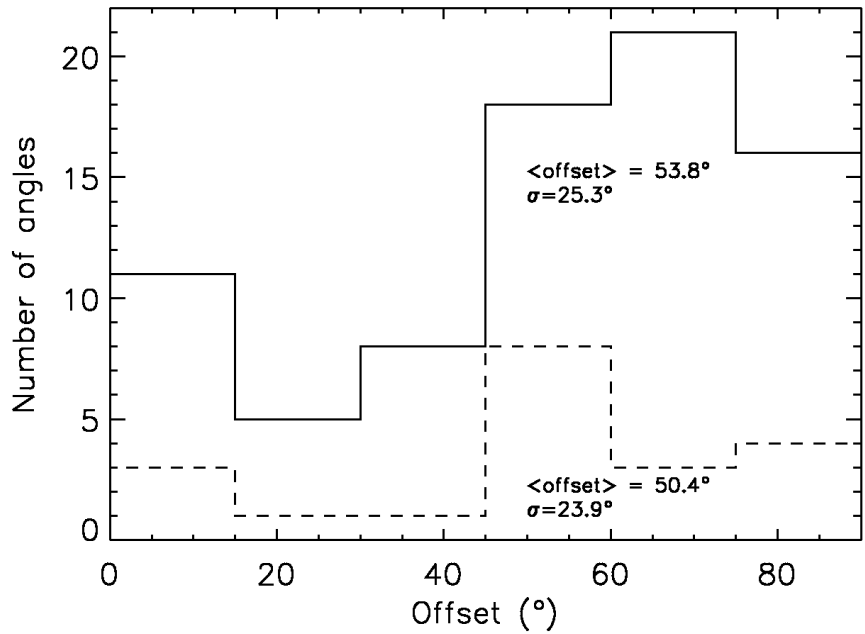}
\caption{Solid line: distribution of angular offsets on the POS between H$_{2}$ jets and CO 
outflows P.A.s with 850 $\mu$m P.A.s of polarization vectors; see Figure \ref{jo}. Dashed line:  
distribution of the mean angular offsets shown in Table \ref{JETS}. Details are given in section \ref{jetsout}.
\label{histo_jo}}
\end{figure}

\begin{figure}
\epsscale{0.8}
\plotone{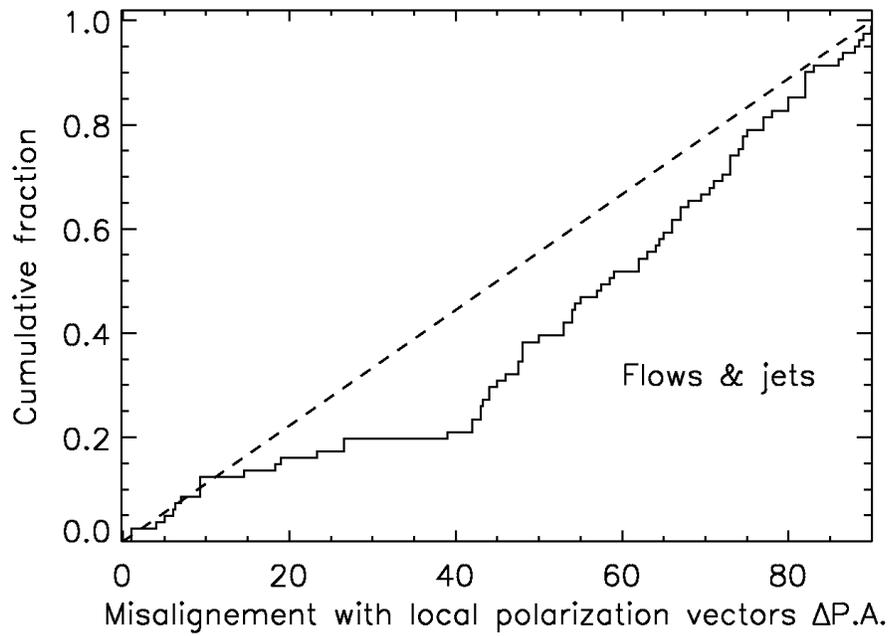}
\caption{Cumulative distribution function of the difference in P.A.s
between local submm polarization vectors and jets/flows. The dashed line is 
the function expected for an infinite randomly oriented sample.   
\label{ks}}
\end{figure}

A histogram of the angular offsets between each polarization position angle and jet/outflow
orientation on the POS can be seen in Figure \ref{histo_jo}.
A histogram of the mean angular offsets from Table \ref{JETS} is also shown by a dashed line. 
With only twenty mean offset estimates, it is difficult to argue that the 
distribution is random rather than normal. So to make a first test,  
the cumulative distribution function (CDF) of the 161 angular offsets
is shown in Figure \ref{ks}.
The straight dotted-line is the CDF expected for an infinite perfectly
random sample. The distribution of the complete flow sample follows
this line reasonably closely. With this sample we find that, statistically, 
the null hypothesis from a Kolmogorov test is accepted, meaning that the observed distribution 
is compatible with a random distribution and that the group of 
flows we compiled could be randomly oriented with respect to the polarization 
pattern of OMC-2 and OMC-3. 
Similarly, a $\chi^2$ test applied to the observed distribution compared to the random distribution  
shows with great probability that these two distributions are similar. 

\section{DISCUSSION} \label{DISCU}

\subsection{A Possible Impact of Jets and Outflows on Submm Polarization Patterns} \label{joimpact}

Based on the statistical tests presented in section \ref{jetsout},
we conclude that no correlation is evident between the 
relative orientation of jets or outflows and polarization vectors on the POS. 
This conclusion is similar to that of \citet{men04} 
in the Taurus molecular cloud complex, an active star-forming region containing no massive stars.
In their study, using absorption polarimetry data, these authors show that T Tauri stars
as a group are apparently oriented randomly with respect to the local magnetic field. This 
indicates that even if the cloud's magnetic field is dominant at large scale, 
its influence largely decreases on the much smaller scale of individual objects
unless the orientation of these individual objects has changed since birth. 
The situation seems to be the same in the OMC-3/2 regions. 
This scenario also implies that if turbulence generated by the outflows is able to misalign
grains in the envelopes of the cores, this effect can not be seen on the POS 
with the resolution of the JCMT. 
For a distance $d \approx 414$ pc, this means that, independently of their orientation 
relative to the line of sight, the energy injected by CO ouflows and H$_{2}$ jets into 
the clouds appears not to have any impact on the polarization patterns which are
observed with the presence of jets and outflows on scales of $\approx 7700$ AU.

In addition, detections of polarized CO emission are now available.
A good introduction about the subject and a summary of some results is given by \citet{for08}. 
Some detections were established by \citet{gir99} in  NGC 1333 IRAS 4A
\citep[see also][]{gir06}, and by \citet{gre99} toward the Galactic center
and in the molecular clouds S140 and DR21.
When the optical depth, $\tau$, and the spatial distribution of the gas 
and of the magnetic fields are favorable, it is possible to detect CO polarization 
vectors perpendicular or parallel to the magnetic field \citep[see][]{kyl83}.
Except in the ``2 pc ring'' where the optical depth $\tau$ is relatively high, 
positive detections found in the other regions are consistent with orientations of 
magnetic fields inferred by submm dust polarimetry. 
Thus, at the scales of their spatial resolution, 
these works suggest that the energy injected by jets and flows should 
have no substantial impact on the net polarization produced by dust grains aligned in the clouds. 
This appears to be consistent with the OMC-3 north region where 
the well-defined polarization pattern suggests that 
whatever the inclination angle relative to the line of sight of the jets and flows,
as well as their orientation on the POS,
the energy injection rate has no influence on the alignment of dust grains seen 
at the scale of the observations. In MMS7 the outflow  
is oriented perpendicular to the 850 $\mu$m polarization vectors covering this area. 
This means that if the situation is the same as in NGC 1333 IRAS 4A
\citep{gir99}, polarized CO emission vectors should be seen perpendicular to these vectors.
In MMS8, 9 \& 10, the analysis of the situation is different since in this part of the sky
the polarization pattern orientation may be due to another superimposed crossing filament
or by a bend of the filament \citep[see][]{mwf01}. 
Since the association of jets and outflows
with their probable progenitors is not always certain because of overlapping effects on 
the sky, a better understanding of the magnetic field structure producing 
the submm polarization pattern observed in the southern part 
of OMC-3 and in the northern part of OMC-2 added to polarized CO emission observations 
could help to constrain these associations.    

The analysis of the situation in OMC-2 is not exactly the same since some of the high intensity regions 
are devoid of detections. 
However, in regions where polarization vectors have a S/N ratio such that $p/\sigma_{p}>3$ 
the general conclusion about the relative orientation of the jets/outflows 
with the submm polarization vectors discussed above is still valid. 

\subsection{Turbulent Angular Dispersions}  \label{TAD}

The second-order structure function of the polarization angles is defined as the average of the squared difference 
between the polarization angle measured at two points separated by a distance $l$ \citep[see equation (5) given by][]{falceta08}. 
Once applied to a grid of pixels containing polarization position angle information, a fit to the square root of this 
function, namely, the Angular Dispersion Function (ADF), gives a method to estimate the turbulent contribution to the total
angular dispersion. First applications of the method and results obtained in regions OMC-1, DR21 Main and M17 are
given by \citet{hil09}. 

\begin{figure}
\epsscale{0.8}
\plotone{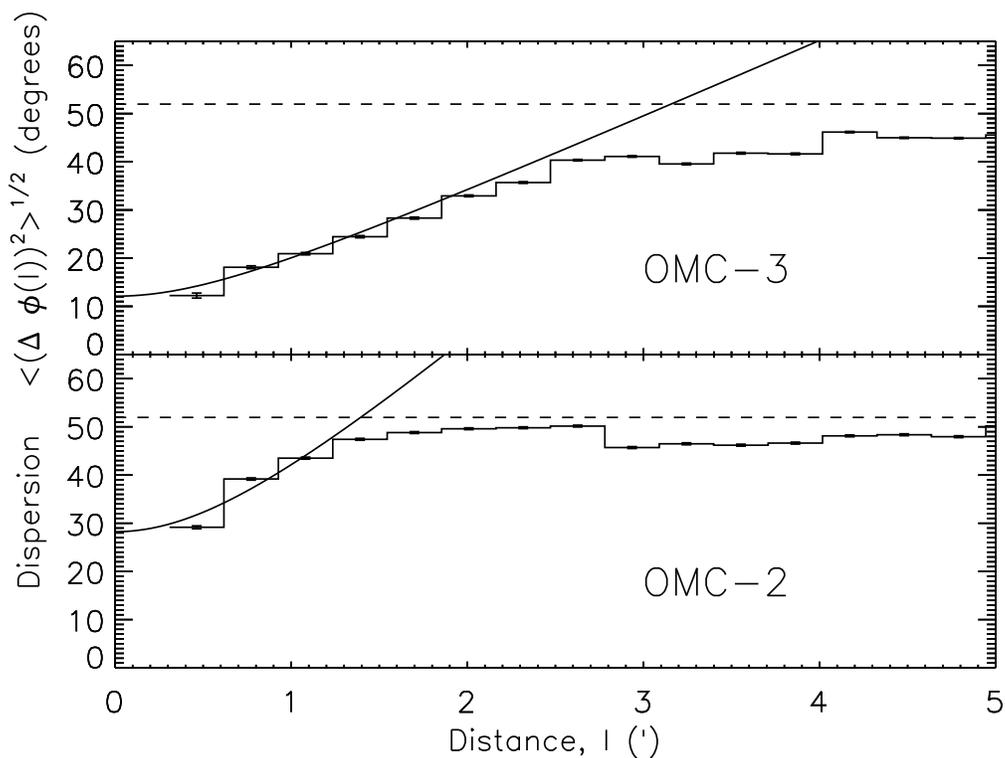}
\caption{ADF, $<\Delta \Phi^2(l)>^{1/2}$, for OMC-3 (top) and OMC-2 (bottom). 
The turbulent contribution to the total angular 
dispersion is determined by the zero intercept of the fit, $b$, to the data at $l=0$ (see Table \ref{subregions}). 
The higher value of $b$ for OMC-2 shows that it is more turbulent than OMC-3.
The dashed-line is the maximum dispersion that would be obtained in case of a purely random 
polarization angle distribution \citep[][]{ser62}. 
\label{DM2OMC23}}
\end{figure}

\begin{figure}
\epsscale{0.8}
\plotone{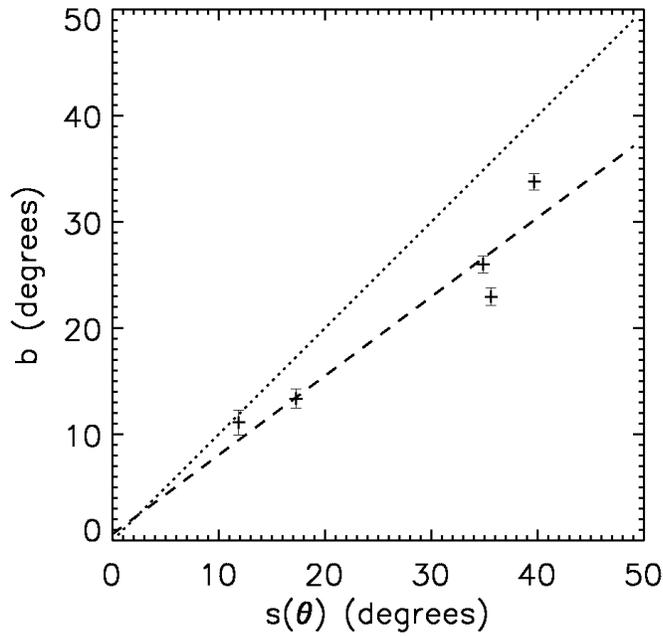}
\caption{Turbulent dispersion component, $b$, estimated with ADF, $<\Delta \Phi^2(l)>^{1/2}$, versus dispersion 
of polarization angles, $s(\theta)$, for regions displayed in Table \ref{subregions}.
The dotted-line expresses equality between these two quantities. The linear fit applied to the data 
and shown by the dashed-line gives the relation $b = (0.60 \pm 1.24) \rm deg + (0.75 \pm 0.04) \times s(\theta)$.
\label{bvsdisp}}
\end{figure}

Figure \ref{DM2OMC23} shows the ADF obtained after application of the method on the 18.6$\arcsec$ 
square pixel grid to the maps of OMC-3 and OMC-2 shown in Figure \ref{polmap}. 
The dashed line shows the maximum dispersion value, $b_{\rm max}=\frac{180^{\circ}}{\sqrt[]{12}}$,
that would be reached in the case of a purely random polarization angle distribution \citep[see][]{ser62}.  
All the maps obtained in several regions of the filament are not shown 
here but the turbulent angular dispersion fitting parameter, $b$, was estimated for each region 
displayed in Table \ref{subregions}. Estimates of $b$ obtained with the correlation method 
for OMC-2/OMC-3, OMC-3, OMC-2 and the five subregions selected along the filament 
(see discussion in section \ref{hole}) are displayed in column (5) of Table \ref{subregions}. 
These values can be compared to the angular dispersion values, $s(\theta)$ obtained with the square root of the 
variance of the data sets displayed in column (4) of the Table. Figure \ref{bvsdisp} shows the variation of $b$ with $s(\theta)$
for the five subregions. The dotted-line expresses equality between these two quantities. A linear fit applied to the data 
and shown by the dashed-line returns the relation, $b = (0.60 \pm 1.24) \rm deg + (0.75 \pm 0.04) \times s(\theta)$, meaning that the 
angular dispersion values estimated with the correlation method are statistically about $25\%$ lower 
than the ones calculated about the mean polarization position angle.

The ratio of turbulent to large scale magnetic field is estimated by using equation (7) given by \citet{hil09}: 
\begin{equation} \label{equ7}       
\frac{<B_{t}^{2}>^{1/2}}{B_{0}}=\frac{b}{\sqrt[]{2-b^2}}
\end{equation}
and values are displayed in column (6) of Table \ref{subregions}. 
In cases where the turbulent component of the field is very small compared to the nonturbulent component, $B_{t} \ll B_{0}$,
\citet{hil09} show that the uniform component of the field can be approximated by the following equation:
\begin{equation} \label{equ8}       
B_{0} \simeq \sqrt[]{8 \pi \rho} \frac{\sigma_{v}}{b}
\end{equation}
We use this approach by assuming a density of $10^{4} \rm cm^{-3}$ and a mean molecular weight of 2.3. Estimates of 
$B_{0}$ are given as a function of $\sigma_{v}$ in the last column of Table \ref{subregions} except for regions MMS1 to MMS7 and 
FIR3 to FIR5 since line width measurements from H$^{13}$CO$^{+}$ $J = 3 \to 2$ were made in OMC-3 MMS6 and OMC-2 FIR4
by \citet{hou00}. We use the values shown in their Table 1 to directly make estimates of $B_{0}$. The line width measurements  
are reported in column (7) of Table \ref{subregions}.
Given the smooth and well defined polarization patterns observed in regions 
MMS1 to MMS7 and MMS8 to MMS10 we are confident that the method is suitable to estimate, within a factor of a few, 
the global mean field component. In the three other subregions, the higher dispersion polarization patterns
suggest that the relation $B_{t} \ll B_{0}$ may not be as well satisfied.
However, all in all, the results shown in Table \ref{subregions} are a first step to intercompare the regions.
 
The estimates of the turbulent angular dispersion components, $b$, displayed in columns (5) of Table \ref{subregions}  
for OMC-3 and OMC-2 are all higher than those estimated by \citet{hil09} in their Table 1. 
We point out that the turbulent component dispersion estimates 
obtained in OMC-3 are of the same order than those found in regions OMC-1 and M17. 
To the contrary, the high value obtained in OMC-2 distinguishes this region from OMC-3, OMC-1, M17 and DR21(Main).
\citet{hil09} find that the dispersions obtained about the mean field orientation are about a factor of 3 
times higher than the ones estimated by using the ADF fitting method. Doing the same comparisons we find 
a factor of about 2 to 1.5 for regions OMC-3/OMC-2, OMC-2 and OMC-3.

Abundances of molecular species and clumping studied by \citet{bat83}, \citet{cas95} and \citet{chi97}
suggest an evolutionary effect from north to south along OMC-2/3. The two regions could have different ages
and OMC-3 could be younger than OMC-2. This could explain why OMC-2 and OMC-3 are so different from the 
point of view of submm polarization data.
On the other hand, the situation is not so clear since Takahashi et al. (2008) show that some intermediate mass objects 
can be at a more evolved stage in OMC-3 than in OMC-2. 
Another explanation could invoke some effects of the radiation field. 
Since OMC-3 is at an higher distance to OMC-1 than is OMC-2, the local ISRF dominated by the bright stars 
located in the Nebulae in front of OMC-1 could be more efficient to shape the OMC-2 region than the OMC-3 region. 
The two regions, OMC-2 and OMC-3, could have the same age but the erosion produced by the local ISRF 
could be stronger on OMC-2 than on OMC-3.
  
\subsection{Ordered vs Turbulent Magnetic Field Components along OMC-2 and OMC-3} \label{compar}

\subsubsection{Depolarization and Turbulence along the Filaments} \label{depol}

\begin{figure}
\epsscale{0.8}
\plotone{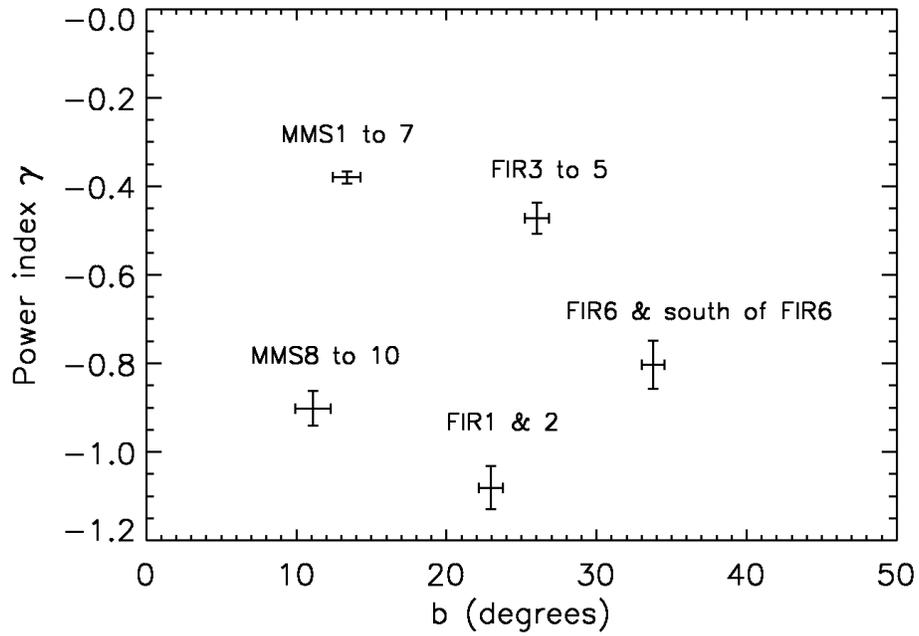}
\caption{Distribution of, $b$, fitting parameter of the turbulent dispersion component estimated with ADF, 
$<\Delta \Phi^2(l)>^{1/2}$, versus, $\gamma$, power index translating the decrease of $p$ with normalized fluxes
(see Figure \ref{pvsi}) for regions displayed in Table \ref{subregions}. 
\label{bvsslopes}}
\end{figure}

\begin{figure}
\epsscale{0.8}
\plotone{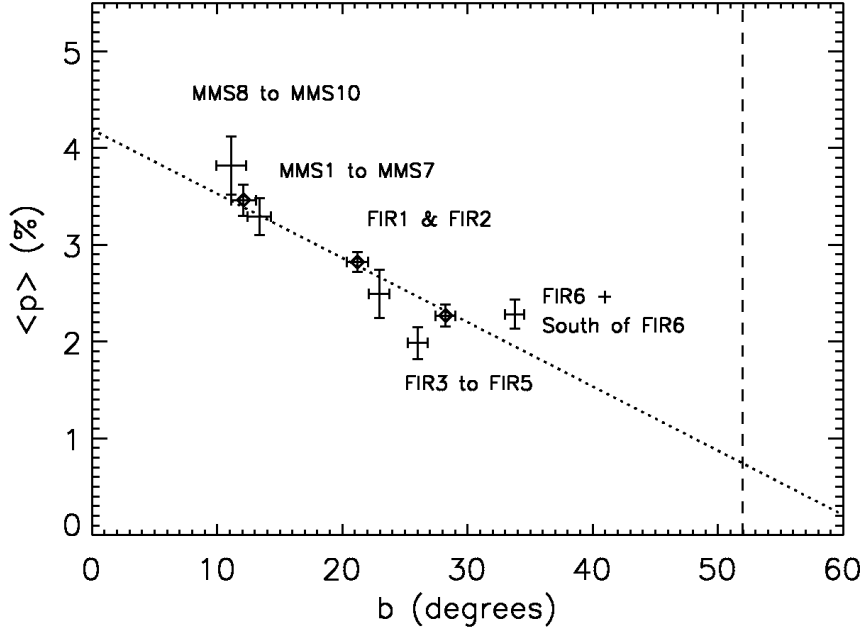}
\caption{Distribution of, $<p>$, mean polarization degree for the regions displayed in Table \ref{subregions}, versus
, $b$, fitting parameter and turbulent dispersion component estimated with the angular dispersion function, 
$<\Delta \Phi^2(l)>^{1/2}$.
Regions OMC-2/OMC-3, OMC-3 and OMC-2 are shown with diamonds. 
The errors on the mean polarization values are shown as vertical error bars. 
The fit discussed in section \ref{depol} is shown by the dotted line. 
The vertical dashed-line displays the maximum turbulent component of about 52$^{\circ}$ theoretically reachable.  
\label{pmeanvsb}}
\end{figure}

Table \ref{subregions} suggests that depolarization is present toward the OMC-2 sources as much as toward OMC-3.
To understand how the decrease 
of polarization with the increase of intensity could be related to turbulent arguments, we show in 
Figure \ref{bvsslopes} the variations of the power indices of the $p-I$ relation, $\gamma$, with the 
turbulent angular dispersion components, $b$, for the five subregions in Table \ref{subregions}.
We find no specific correlation meaning that the decrease of the polarization degree observed 
appears independent of turbulent flows or turbulent effects that could be 
present into the densest regions of the cloud. On the other hand, the variation of the 
the mean polarization degrees, $<p>$, with the turbulent angular 
dispersion estimates, $b$, of the regions considered in Table 
\ref{subregions} and shown in Figure \ref{pmeanvsb} suggests an anti-correlation between the two parameters. 
If this trend is real, it would mean that, independently of the ordered magnetic field structure component, 
the mean polarization degree observed in a given region could be a function of the turbulent magnetic field component. 
To test this hypothesis we did a linear fit of the form $ <p> = <p_{0}> + c_{1} \times b$, to the eight points 
displayed in Figure \ref{pmeanvsb}. The results are $<p_{0}>=4.2 \% \pm 0.2\%$ and 
$c_{1}=-0.07 \% \rm deg^{-1}  \pm 0.01 \% \rm deg^{-1}$, 
where the fit is constrained by taking into account the errors on $p$. 
The errors are estimated by dividing the standard deviation of each data set by the square root of the number
of data. In addition, a fit of the variations 
of $b$ with $<p>$ (not shown here) of the form $b = b_{0} + c_{2} \times <p> $, taking into account the uncertainties on $b$, 
gives the results $b_{0}=54.2^{\circ} \pm 1.5^{\circ}$ and $c_{2} = -11.8 \rm deg \%^{-1} \pm 0.6 \rm deg \%^{-1}$.
Given the errors on $p$, the first fit constrains reasonably the variations of $<p>$ with $b$.
Within the uncertainties the intersect of the fit with the abscissa axis gives a value of $b_{0}$ which 
is hardly consistent with the 
expected value, $b_{\rm max}$ of about $52^{\circ}$, that would be found in the case of a purely random 
polarization position angle distribution \citep[][]{ser62}.
On the contrary, the second fit returns a value $b_{0}$ relatively close to that of $b_{\rm max}$ 
suggesting that if such a region was observable the mean polarization degree could be about zero. 

To test if the trend observed in OMC-2/OMC-3 is consistent with the polarization properties of OMC-1 
we use the turbulent dispersion parameters, $b$, estimated by \citet{hil09} in their Table 1 
and Figure 3 of \citet{vai08} where median polarization ratios are displayed for different wavelengths and for 
several regions. The mean polarization percentage observed at 350 $\mu$m by \citet{hou04} is $<p> = 2.7 \%$
with a dispersion $\sigma_{p}=1.4\%$. 
Using Figure 3 shown by \citet{vai08} and a wavelength ratio of 1.3, a mean polarization of about 3.5 $\%$ 
should be representative of the mean degree of polarization observed at 850 $\mu$m in OMC-1 with 12$^{\arcsec}$
resolution. Neglecting the effects that could produce a slightly lower resolution of 14$^{\arcsec}$ and with 
the assumption that the turbulent ratios derived with 350 $\mu$m and 850 $\mu$m polarization maps would be similar  
this yields a point $(b=8.3\pm0.3, <p> \approx 3.5 \%)$ that appears to be consistent with our fit.

A similar trend was observed along the Pipe Nebula by \citet{alv08} with $R$-band visible polarization data
collected for about 12000 stars. In this case the dispersion of several subsets of polarization position angles is used 
to estimate the turbulence of regions having a mean density $n(\rm H_{2}) \approx 10^{3}$ cm$^{-3}$
and is compared to the mean polarization degrees of the subsets of data. The regions probed in the 
Pipe Nebula have densities about an order of magnitude lower than the density considered in our 
calculations where submm data would mostly probe the inner fields embedded in the filaments.
 
The estimates of the mean field strength displayed in Table \ref{subregions} for regions MMS1 to MMS7 and 
FIR3 to FIR5 are of the same order. This result dismisses the hypothesis that the dust grain alignment efficiency 
could decrease with the magnetic field strength.
In addition, mean inclination angles of the magnetic field from the LOS were estimated by \citet{hou04}
to vary from about $73^{\circ}$ to about $80^{\circ}$ in three dense regions displayed along OMC-2/OMC-3.  
If these inclination angles are representative of the mean inclination angle of the magnetic field 
along the filaments and if the grains are aligned with roughly the same efficiency everywhere then 
the degree of polarization would not be very sensitive to this parameter. 
These two points do not dismiss the possibility that the mean polarization degree of some 
regions could be a function of the level of turbulence in the region, but nor do they rule out 
the possibility of complex structures of an ordered magnetic field component. In cases where the 
large scale magnetic field dominates over the turbulence it could be that some
superposition effects along the LOSs would produce a decrease of the polarization 
degree with an increase of the polarization position angle dispersion over some regions. 
These aspects are discussed in the following sections.

\subsubsection{Effects of Steady-State Magnetic Field Models} \label{steadystate}

The Turbulent-to-Mean magnetic field strengths ratios displayed in Table \ref{subregions} imply
intensities of the turbulent magnetic component about 6 to 7 times lower than the mean magnetic field component
into the OMC-3 filaments. This ratio supports the steady state magnetic field approach proposed by \citet{fp00d}
and applied to the OMC-3 region as discussed by \citet{mwf01}. In the work of \citet{fp00d} the maximum polarization 
degree is calibrated with submm observations and dust grains are aligned perpendicular to the magnetic fields.
The combination of toroidal with poloidal magnetic fields components 
leads to superposition effects along the same line-of-sight. This geometrical effect decreases the net 
polarization on the POS and renders uniform magnetic fields indistinguishable from a helical field geometry.
The model can reproduce the $r^{-2}$ density profile\footnote{Here $r$ is the radial distance, in cylindrical coordinates.} 
observed by \citet{joh99} and could explain the 
depolarization observed along the spine of the filament. A similar approach was followed by \citet{gon05} with a focus
on molecular cloud cores and shows that tangling of the magnetic field orientations by the effects of gravity could produce 
polarization maps showing an significant angular dispersion from the mean uniform field.  
Since it is not clear that OMC-2 is a clear case of a magnetically dominated region with respect to the turbulence, it 
would be interesting to test under which conditions steady-state magnetic models could 
reproduce the polarization properties observed in this region.

\subsubsection{Variation of MHD models}

The statistical results shown in Figure \ref{pmeanvsb} discussed above in section \ref{depol} 
suggest that turbulence could be a parameter regulating the mean degree of polarization observed in a given 
polarization map. One open question would be to understand to what extent MHD models could reproduce 
such a decrease of polarization.

\citet{falceta08} present results for turbulent, isothermal, three-dimensional  
simulations of sub/supersonic and sub/super-Alfv\'enic 
cases. Dust grains are assumed perfectly aligned and perpendicular to the magnetic field. 
Depolarization is due to the dispersion increase of the polarization position angles and is a function of the MHD 
regime considered as well as of the orientation of the initial uniform magnetic field with respect to the LOS. 
The dust total intensity is assumed to be proportional to the column density. 
The high resolution of the simulations provides less homogeneous magnetic field structures and higher density contrasts
than previous models. 

On the other hand, \citet{cho05} show that under peculiar conditions depolarization could occur if grains embedded in dark clouds
are aligned by radiative torques (RATs) such that their long axis is perpendicular to the magnetic field.
In their model, the nonturbulent field is about 2 times stronger than the fluctuating magnetic field, a condition encountered in 
OMC-2 (see Table \ref{subregions}) and the ordered component is assumed to be uniform and in the POS. This condition may exist
along the length of the OMC-2 and OMC-3 filaments \citep[][]{hou04}. Additionally, the results proposed 
by \citet{cho05} are valid for clouds without embedded massive stars, a condition valid in OMC-2/3 where 
forming stars are of intermediate \citep[see][]{tak08}, or lower mass. 
Two interesting extensions to the \citet{cho05} work
were proposed by \citet{pel07} and by \citet{bet07} but the analysis of the simulations is focused on 
the effects of RATs on depolarization rather than on the effects of the turbulent regimes on depolarization.
Alignment by RATs is not considered by \citet{falceta08} but an anti-correlation between the polarization degree 
and the column density, with exponent $\gamma \sim -0.5$ is predicted, due to random cancellation of polarization 
vectors along the LOS. This value is close to some values obtained on larger scales in OMC-2 and OMC-3 under 
conditions which are discussed in more detail in the following section.
Comparisons done by \citet{falceta08} between four different MHD regimes show a degeneracy between 
the Alfv\'enic Mach number and the angle between the mean magnetic field and the LOS. They discuss the effects of 
different resolutions on the structure function of the polarization angle and the applicability of these 
structure functions to the determination of turbulent cut-off scales. 

A comparison of the $14^{\arcsec}$ resolution 850 $\mu$m  
P.A. histogram displayed in Figure \ref{anghistot} with the results of Figure 4 of \citet{falceta08} would reject the presence of 
super-Alfv\'enic modes in OMC-3 and probably in OMC-2 as well. This inference would be consistent with the 
lack of correlation between jets/outflows and polarization vectors discussed in section \ref{joimpact}. But, 
as suggested in other regions by \citet{pad04}, 
this would not rule out the presence of super-Alfv\'enic modes at smaller scale, in cores of about 0.25 pc 
in size in the OMC-2/3 regions for a distance to the clouds of about $d = 414 $ pc \citep[see][]{men07}. 
Finally, because of the resolution of our data, a direct comparison of our results with the 
structure function of the polarization position angle calculated for different MHD regimes is not done.    

\subsubsection{Variations of Power Index with Column Density Structure} \label{mhdrats}

Figure \ref{pmeanvsb} illustrates that the mean polarization degree of any region could be controlled by the MHD regime 
but it does not give any indication about the possible effect of the MHD regime on the level of depolarization.  
Figure \ref{bvsslopes} shows an absence of correlation between the power index of the $p-I$ relation and the 
turbulence parameter, $b$, however, and suggests that one or more
phenomena other than turbulence should be considered. 
To inform this discussion, we consider possible variations in column density structure and the power law index, 
our best quantifier of depolarization.

\begin{figure}
\epsscale{0.8}
\plotone{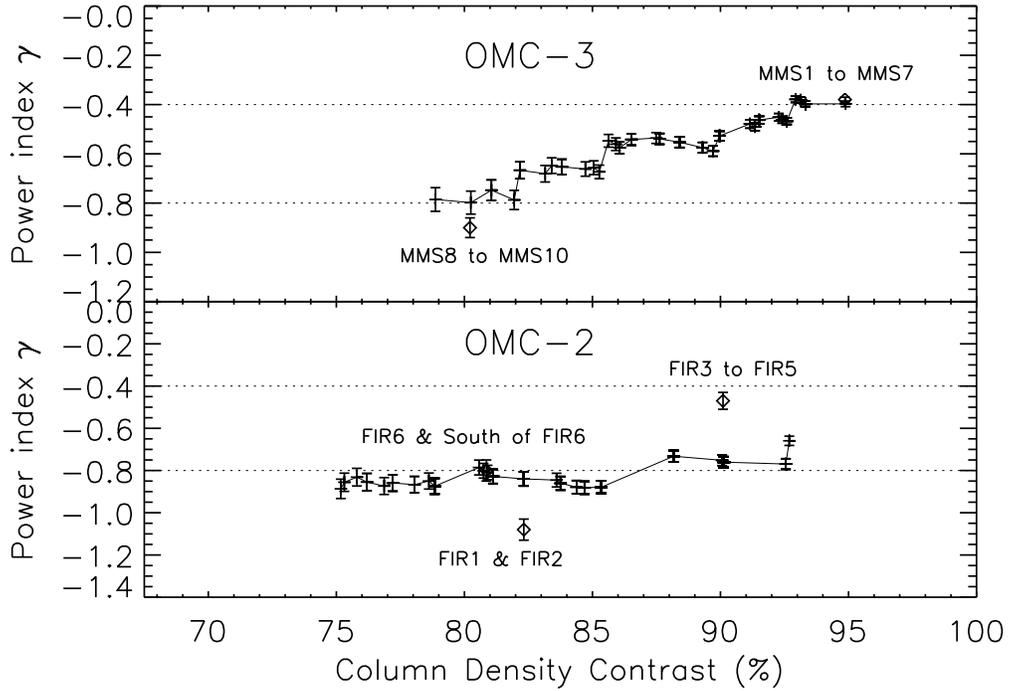}
\caption{Distribution of the power index of the $p-I$ relation with the Column Density Contrast, 
$CDC=(Flux_{\rm max} - Flux_{\rm min})/Flux_{\rm max}$.
Diamonds show the $CDC$ and power index values for the subregions identified in Table \ref{subregions}, 
without masking any values. 
The solid lines trace the behavior of the power index as the $CDC$ varies with each subsequent masking 
of high intensity pixels (for a full explanation, see the text). 
The OMC-3 region is shown in the top panel and the OMC-2 region in the bottom panel.
The horizontal dashed lines show the two regimes discussed in section \ref{mhdrats}. 
\label{slopesvsdc}}
\end{figure}

Estimates of the power indices of the five subregions from Table \ref{subregions} are shown as a function of the 
column density contrast, $CDC=(Flux_{\rm max} - Flux_{\rm min})/Flux_{\rm max}$, in Figure \ref{slopesvsdc}. 
On a statistical basis, the variation of the power index with the $CDC$ is estimated by considering 
the several maps obtained by masking high intensity pixels above specific cutoffs. 
These variations of the power indices with the $CDC$s are shown by the solid lines in Figure \ref{slopesvsdc}. 
The highest $CDC$ values are derived from our original maps (Figure \ref{polmap}).  To establish the power index of lower column 
material, for which we assume lower fluxes to be a proxy, we methodically masked the highest pixels by using a step 
of 33 $mJy$/18.6$^{\arcsec}$ in column density and then recalculated the $CDC$ and power index of the resulting $p-I$ relation. 
By repeating this process, we are able to assess the impact of lower and lower column material on $\gamma$. 
We stopped at the level of $26\%$ of the peak in OMC-3 (71 of 116 pixels were used in the process) and 30$\%$ of the peak in OMC-2
(108 of 135 pixels). Below these values, divergences were observed in the estimates of $\gamma$, 
likely due to the small sampling statistics. 
 
In OMC-3, the power index shows almost a linear decrease from $\gamma=-0.8$ down to $\gamma=-0.4$.
The upper value suggests a first regime reflecting the statistical level of depolarization obtained when the 
cores and the high density flux regions are included and well represented. The lower value suggests another regime 
corresponding to the statistical level of depolarization obtained when the high density flux regions are avoided.
Along OMC-2, only a small range of variation is observed with $\gamma \approx -0.8$, perhaps reflecting the fact 
that polarization was not detected in many of the brightest regions.
The two regimes are illustrated by the two horizontal dashed lines shown in Figure \ref{slopesvsdc}.  

\begin{figure}
\epsscale{0.8}
\plotone{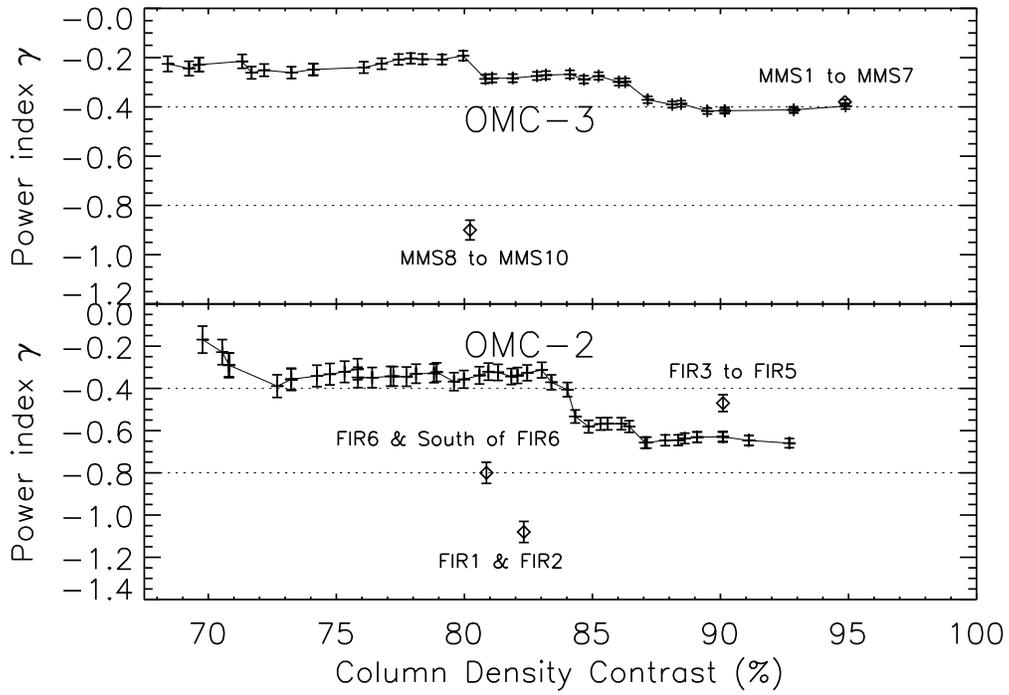}
\caption{Same as Figure 14, but instead of high intensity pixel masking, 
low intensity pixels are progressively masked out.
\label{slopesvsdcup}}
\end{figure}

To test the possibility of two regimes, we applied the same process in reverse, preferentially masking the 
lowest flux pixels until reaching 31$\%$ of the peak in OMC-3 (31 of 116 pixels were used in the process)
and 30$\%$ of the peak in OMC-2 (25 of 135 pixels). The subsequent variations of the 
power index of the $p-I$ relation with the $CDC$ are plotted on Figure \ref{slopesvsdcup}. 
Here again the effect of the cores seems to be under represented on larger scale in OMC-2 and the power 
index is constant and about -0.6 in the density contrast range $CDC$ $\approx 84 \%$ to $93 \%$. Once the lowest
density structures are avoided, however, a second regime appears and the power index is constant and about 
-0.4 in the density contrast range $CDC$ $\approx 73 \%$ to $84 \%$. 
In OMC-3 the power index is constant and about -0.4 in the density contrast range $CDC$ $\approx 87 \%$ to $95 \%$.
Below a $CDC$ $\approx 87 \%$ the power index increases from about -0.4 to -0.2
showing the effects a reduction of the coverage mapping centered around high density regions could produce. 

Finally, when comparisons are done between the five subregions, the effect of the presence of the cores 
on the measured power index is clearer.
The two regions FIR3 to FIR5 and MMS1 to MMS7 are consistent with the first regime where the cores are 
included since the maximum intensity pixel used to normalize the intensity of the $p-I$ relation
are the same as those used in OMC-2 and OMC-3, respectively. Interestingly, the relatively low density
region FIR6 and South of FIR6 appears consistent with the second regime where the effects of the cores are avoided. 
Departures from the two regimes are observed in the MMS8 to MMS10 and particularly FIR1 \& FIR2 regions. 
The maximum intensity pixels used to normalize the intensity of 
the $p-I$ relation are small in these regions compared to the one used in region FIR6 and South to FIR6.
This makes the sizes of the samples of regions MMS8 to MMS10 and FIR1 \& FIR2 too small to be representative of 
the density structure of one of the two apparent regimes observed on larger scale.

In conclusion, if representative of the column density structure of the molecular cloud, the presence of cores can
lead to two distinct regimes of depolarization: one with a shallower power index reflecting the cores 
and a steeper value representative of lower column density material. 
We note, as pointed out in observational works \citep[see][]{and07,whit08}, that dust alignment by RATs could be a promising 
ingredient for understanding depolarization. In this framework, a possible explanation for the 
shallower power index in cores could be the growth of larger, non-spherical grains in dense, cold condensations.
As a consequence the upper cutoff
of the power-law distribution of grain sizes will be higher in maps including cores than without cores
and the power index of the $p-I$ relation be lower \citep[see Figure 6 in the work of][]{cho05}. 
An alternative explanation could be an enhancement of gravity high enough to 
distort the magnetic field orientations in the cores. This mechanism will tend to decrease the net polarization
and the values of the power index will change accordingly \citep[see][and details given in section \ref{steadystate}]{gon05}.  
We note that our analysis could be subject to a bias if 
the resolution of the instrument is too small to properly sample the column density structure of the clouds. 
Because the \citet{wil03} interferometric mapping in OMC-2/3 precludes any multi-core scenario, the analysis 
of our data should not be subject to a bias due to intercept of cores along the same LOS, a possibility
discussed by \citet{pel07}.

\section{SUMMARY} \label{SUM}

850 $\mu$m SCUBA polarization data of OMC-2 were homogeneously reduced in combination with archived data of OMC-3.
Our main results are:

1. In OMC-2, the polarization pattern shows more variations of polarization 
position angles on spatial scales similar to those of the two well-ordered polarization 
patterns observed in OMC-3. The mean degree of polarization is lower in OMC-2 than in OMC-3.
We find that vectors in FIR1 suggest a realignment of the polarization vectors with the 
filament in the region south of OMC-3.

2. In regions of existing 350 $\mu$m polarization data, we find that 
except in FIR6, 850 $\mu$m polarization vector patterns are similar to those at 350 $\mu$m. 
Significant detections are only available at 850 $\mu$m in and around three regions, 
namely: MMS7, FIR1 \& FIR2, and in the region South of FIR6. 

3. A comparison of the offsets between P.A.s of CO outflows/H$_{2}$ jets and polarization 
vectors suggests that no specific orientation of these outflows relative to the 
polarization patterns can be found. This fact suggests that if dust grains are 
generally aligned with their long axis perpendicular to the magnetic field, there 
is no correlation between outflows and the mean 
magnetic field orientation on the POS, at least to the $14^{\arcsec}$ beam resolution of our data. 

4. Based on the hypothesis that turbulence is present along the filaments, second order
structure functions of the polarization position angle show that OMC-3 is a less turbulent region than
OMC-2. OMC-3 appears to be a clear case of a magnetically dominated region with respect to the turbulence.
However for OMC-2, it is not clear that this is the case.
In OMC-2 and OMC-3 the dispersions obtained about the mean field orientation are factors of
1.5 to 2 times higher than those estimated with the angular dispersion fitting method.
Estimates of the projected magnetic field strengths associated with the objects MMS6 and FIR4
are calculated to be 0.19 mG and 0.13 mG, respectively. 

5. A more in-depth analysis provides estimates of the power index, $\gamma$, also 
known as the depolarization parameter,  
the dispersions obtained about the mean field orientation $s(\theta)$ and with the angular 
dispersion fitting method $b$, the turbulent-to-mean magnetic field strength ratio 
$<B_{t}^{2}>^{1/2}/B_{0}$ for five regions along OMC-3 and OMC-2. 
We find an anti-correlation between $<p>$, the mean polarization degree, and the turbulence parameter $b$, 
meaning that the level of turbulence in a region could regulate the mean polarization degree observed.
No specific correlation is found between $\gamma$ and the turbulence parameter $b$. 

6. When steady state models are considered, two scenarios can explain the polarization at the southern 
edge of OMC-3: a bent filament and a second filament oriented almost orthogonal to the integral shaped filament
\citep[see][]{mwf01}. 
It remains to be quantified how the relative effects of steady state and turbulent magnetic field models 
contribute to the observed polarization in OMC-2.

7. If the mapping sufficiently reflects the column density structure of the clouds, 
a statistical analysis suggests the presence of two depolarization regimes in our maps.  
One regime including the effects of the cores, the other one excluding it. 

The authors would like to thank G. Schieven at the JCMT for his assistance during and after observing, R.L. Curran 
at the Dublin Institute for Advanced Studies for her help during the data reduction process and an anonymous referee
for useful suggestions. The research of F.P. was supported by the Conseil de Recherche en Sciences Naturelles et 
en G\'enie du Canada. This author also thanks the Funda\c{c}\~ao de Amparo \`a Pesquisa do Estado de S\~ao Paulo 
(FAPESP grant number 2007/56302-4) for support during the second phase of the present research. 
The research of P.B. is supported by the Conseil de Recherche en Sciences Naturelles et en G\'enie du Canada. 
The research of B.M. is supported by a Plaskett Fellowship at the National Research Council of Canada.

\clearpage

\clearpage

\clearpage

\clearpage

\clearpage

\clearpage

\clearpage

\clearpage

\clearpage

\clearpage

\clearpage

\clearpage

\clearpage

\clearpage

\clearpage

\clearpage

\clearpage

\begin{deluxetable}{cccccc} 
\tablewidth{0pt}
\tabletypesize{\scriptsize}
\tablecaption{Observational Parameters for Jiggle Mapping Polarimetry in OMC-2. \label{JIG}}
\tablehead{
 \colhead{Pointing Center} & \colhead{Pointing Center} &
 \colhead{Chop Position Angle} & \colhead{Distance} &
 \colhead{Number of} \\
 \colhead{R.A.(J2000)} & \colhead{Dec.(J2000)} &
 \colhead{east from north ($^{\circ}$)} & \colhead{(\arcsec)} &
 \colhead{ observations}
}
\startdata
05 35 22.2&-05 06 40.1&112& 180&9\\
05 35 25.0&-05 07 49.3& 90& 150&12\\
05 35 26.8&-05 09 13.5& 90& 150&12\\
05 35 25.0&-05 10 37.4& 90& 150&18\\
05 35 24.9&-05 12 07.3& 90& 150&15\\
05 35 21.9&-05 13 19.1& 90& 150&12\\
05 35 18.8&-05 14 54.9& 90& 150&12\\
\enddata
\end{deluxetable}

\clearpage

\begin{deluxetable}{lrrrrrrc} 
\tablewidth{0pt}
\tabletypesize{\scriptsize}
\tablecaption{850 $\mu$m polarization data through OMC-2. \label{DATA}}
\tablehead{
  \colhead{ID$^{(a)}$}                  & \colhead{R.A.(J2000)} &
  \colhead{Dec.(J2000)}  & \colhead {$p$} &
  \colhead{$\sigma_{p}$}        & 
  \colhead{$\theta^{(b)}$} &
  \colhead{$\sigma_{\theta}$}           & \colhead{Flux} \\
  \colhead{}                    & \colhead{($^{\rm h}$ $^{\rm mn}$ $^{\rm s}$)} &
  \colhead{($^{\circ}$ $\arcmin$ $\arcsec$)}       & \colhead{($\%$)} &    
  \colhead{($\%$)}              & \colhead{($^{\circ}$)} &
  \colhead{($^{\circ}$)}            & \colhead{$(Jy/18.6^{\arcsec}$ beam)} 
}
\startdata

 1  & 05 35 18.89  &      -05 15 43.9  &            1.34   &       0.39  &       -3.1  &        8.2   & 1.5 \\            
 2  & 05 35 22.61   &     -05 15 25.4   &           4.54  &        0.61  &       -11.5  &       3.8   & 0.9 \\            
 3  & 05 35 21.37  &      -05 15 25.4   &           2.46  &        0.50  &       -6.8  &       5.7   &  1.1\\            
 4  &05 35 20.13   &     -05 15 25.4   &           2.54   &       0.24   &      -32.1  &       2.8    &  2.3\\            
 5  &05 35 17.65   &     -05 15 25.4   &           2.73   &       0.51   &      4.9  &        5.2   &  1.1\\            
 6  &05 35 16.41   &     -05 15 25.4   &           2.33   &       0.58   &       -2.7  &       6.3   &  1.0\\            
 7  &05 35 22.61   &     -05 15 06.9   &           3.83   &       0.46  &       -22.0    &      3.4   &  1.3\\            
 8  &05 35 21.37   &     -05 15 06.9   &           3.50   &       0.35   &      -24.7 &       2.9    &  1.6\\           
 9  &05 35 20.13   &     -05 15 06.9   &           2.78    &       0.29   &      -31.8  &       3.0   &  1.9\\            
 10  &05 35 18.89   &     -05 15 06.9  &            0.99  &       0.32  &       -3.9  &       9.0  &  1.5\\             
 11  &05 35 17.65   &     -05 15 06.9  &            1.52  &        0.42  &       -3.3  &       7.0  &  1.2\\             
 12  &05 35 16.41   &     -05 15 06.9  &            4.18  &        0.45  &       9.7   &       2.9   &  1.2\\            
 13  &05 35 22.61   &     -05 14 48.3  &            2.70  &        0.48  &       -21.0  &       5.0   &  1.5\\            
 14  &05 35 21.37   &     -05 14 48.3  &            1.52  &        0.27   &       -18.7  &       5.0  &  2.1\\             
 15  &05 35 16.41   &     -05 14 48.3  &            2.95  &        0.56  &       -2.4   &       5.3    &  0.9\\            
 16  &05 35 20.13  &      -05 14 29.8  &            2.04  &        0.40  &       -48.2  &       6.0   & 1.2 \\            
 17  &05 35 18.89  &      -05 14 29.8  &            2.34  &        0.58  &       10.0    &      6.8   & 1.0 \\             
 18  &05 35 17.65  &      -05 14 29.8  &            4.16  &        0.68   &      0.2   &      4.4   &  1.0\\            
 19  &05 35 16.41  &      -05 14 29.8  &            3.86  &        0.65  &       -0.3  &      5.3   &  0.7\\            
 20  &05 35 23.86  &      -05 14 11.2  &            2.98  &        0.59  &       0.12   &      5.8  &  1.0\\             
 21  &05 35 22.61  &      -05 14 11.2  &            2.08  &        0.39  &       28.4    &      5.4  &  1.5\\             
 22  &05 35 18.89  &      -05 14 11.2  &            2.07   &        0.45  &       24.7    &      6.1  &  1.2\\             
 23  &05 35 22.61  &      -05 13 52.7  &            4.33  &        0.63  &       24.6    &      4.1  &  0.9\\             
 24  &05 35 21.37  &      -05 13 52.7  &            1.97  &        0.48  &       -71.9    &      6.8   & 1.1 \\            
 25  &05 35 18.89  &      -05 13 52.7  &            3.01  &        0.56  &       26.2   &       5.1   &  0.9\\            
 26  &05 35 17.65  &      -05 13 52.7  &            3.21  &        0.59  &       27.9    &      5.2   &  0.9\\            
 27  &05 35 25.10  &      -05 13 34.2  &            2.22  &        0.52  &       88.3     &      6.9   &  0.8\\            
 28  &05 35 23.86  &      -05 13 34.2  &            1.10  &        0.36  &       -80.8   &      8.5  &  1.3\\            
 29  &05 35 22.61  &      -05 13 34.2  &            1.66  &        0.33  &       31.0   &       5.8  &  1.7\\             
 30  &05 35 21.37  &      -05 13 34.2  &            1.10   &       0.31   &      64.3   &       8.1  &  1.9\\             
 31  &05 35 17.65  &      -05 13 34.2  &            2.39   &       0.71   &      50.0   &       8.0  &  0.9\\             
 32  &05 35 25.10  &      -05 13 15.6  &            1.87  &        0.33  &       4.7   &       5.1  &  1.2\\             
 33  &05 35 17.65  &      -05 13 15.6   &           2.37   &       0.71  &       25.4   &       8.3  &  0.9\\             
 34  &05 35 26.34  &      -05 12 57.1  &            2.04   &       0.52   &       8.6     &       7.4   &  0.8\\             
 35  &05 35 25.10  &      -05 12 57.1  &            0.96  &       0.27   &       35.6   &       7.6  &  1.5\\             
 36 & 05 35 23.86  &      -05 12 57.1  &            1.27    &       0.15  &       32.8   &       3.6  & 2.4 \\             
 37 & 05 35 22.61  &      -05 12 57.1  &            0.91  &       0.11   &       70.6    &       3.4  &  3.6\\             
 38 & 05 35 21.37  &      -05 12 57.1   &           0.78  &       0.14  &       42.2   &       5.0  &  3.0\\             
 39 & 05 35 20.13  &      -05 12 57.1  &            0.99  &       0.28  &       3.9   &       8.0   &  1.6\\            
 40 & 05 35 18.89  &      -05 12 57.1  &            3.87   &       0.54  &       44.0   &       3.9  &  1.0\\             
 41 & 05 35 26.34  &      -05 12 38.5  &            1.56   &       0.42  &       -66.8  &       7.3   &  1.2\\            
 42 & 05 35 25.10  &      -05 12 38.5  &            2.42   &       0.21   &       -9.1  &       2.5  &  2.3\\             
 43 & 05 35 23.86   &     -05 12 38.5  &            0.78  &       0.13  &       -38.0  &       5.0  &  3.5\\             
 44 & 05 35 21.37  &      -05 12 38.5  &            0.73   &       0.24  &       31.9    &       9.2  &  1.9\\             
 45 & 05 35 28.82  &      -05 12 20.0  &            4.94   &       0.62   &       22.1   &       3.6   &  1.0\\            
 46 & 05 35 27.58  &      -05 12 20.0   &           3.05     &       0.53  &       8.7   &       4.6    &  1.1\\            
 47 & 05 35 28.82  &      -05 12 01.5  &            5.17  &        0.72  &       13.4   &       4.1  &  1.0\\             
 48 & 05 35 27.58  &      -05 12 01.5  &            1.57  &        0.43  &       2.8   &       7.6   &  1.1\\             
 49 & 05 35 26.34  &      -05 12 01.5  &            1.22    &       0.38  &       -26.4  &       8.5  &  1.4\\             
 50 & 05 35 23.86  &      -05 12 01.5  &            1.34  &        0.22  &       -80.5   &       4.9   &  2.4\\            
 51 & 05 35 28.82  &      -05 11 42.9  &            2.17  &        0.59   &      -44.4  &       7.2   & 0.9 \\            
 52 & 05 35 27.58  &      -05 11 42.9  &            1.61   &       0.23  &       1.8   &       5.3   &  1.5\\            
 53 & 05 35 25.10  &      -05 11 42.9  &            1.47   &       0.31  &       -51.7  &       6.2    &  1.3\\            
 54 & 05 35 23.86  &      -05 11 42.9  &            1.54   &       0.35  &       -80.4  &       6.0   &  1.5\\            
 55 & 05 35 27.58  &      -05 11 24.4  &            0.97  &        0.26  &       8.0   &       7.7  &  1.7\\             
 56 & 05 35 26.34  &      -05 11 24.4  &            1.14  &        0.36  &       22.2   &       8.7   &  1.2\\            
 57 & 05 35 23.86  &      -05 11 24.4  &            3.01  &        0.45  &       -41.6  &       4.3  &  1.0\\             
 58 & 05 35 30.06  &      -05 11 05.8  &            2.76  &        0.74  &       -9.8  &       7.7  &  0.9\\             
 59 & 05 35 28.82  &      -05 11 05.8  &            1.68  &        0.36  &       -15.5  &       6.1  &  1.4\\             
 60 & 05 35 27.58  &      -05 11 05.8  &            2.19   &        0.33   &      13.2   &       4.2   &  1.6\\            
 61 & 05 35 26.34  &      -05 11 05.8  &            1.68  &        0.33  &       48.0    &       6.5   &  1.4\\            
 62 & 05 35 25.10  &      -05 11 05.8  &            1.92  &        0.50   &      -88.7  &       7.4  &  0.9\\             
 63 & 05 35 23.86  &      -05 11 05.8  &            2.46  &        0.50  &       -64.5  &       5.7  &  0.8\\             
 64 & 05 35 30.06  &      -05 10 47.3  &            2.55  &        0.68  &       -69.2  &       7.4  &  1.0\\             
 65 & 05 35 28.82  &      -05 10 47.3  &            1.72  &        0.41  &       -37.6  &       6.6   &  1.7\\            
 66 & 05 35 26.34  &      -05 10 47.3  &            2.08   &       0.24  &       -82.9  &       3.7   &  2.5\\            
 67 & 05 35 23.86  &      -05 10 47.3  &            2.74  &        0.61  &       -55.5  &       6.4  &  0.9\\             
 68 & 05 35 28.82  &      -05 10 28.8  &            1.64  &        0.26   &       -24.1  &       4.5  &  2.4\\             
 69 & 05 35 27.58  &      -05 10 28.8  &            1.20   &        0.17  &       -58.8  &       3.9  &  3.9\\             
 70 & 05 35 26.34  &      -05 10 28.8  &            2.06  &        0.13  &       -75.5  &       1.9   &  4.8\\             
 71 & 05 35 25.10  &      -05 10 28.8   &           2.02  &        0.23  &       -65.5  &       3.2  &  2.6\\             
 72 & 05 35 23.86  &      -05 10 28.8  &            5.26   &        0.43   &      -58.3  &       2.3  &  1.2\\             
 73 & 05 35 22.61  &      -05 10 28.8  &            4.69  &        0.45   &      -52.9  &       2.8   &  1.2\\             
 74 & 05 35 27.58  &      -05 10 10.2  &            0.56   &       0.13   &       -37.8   &      6.5   &  5.7\\            
 75 & 05 35 26.34  &      -05 10 10.2  &            1.58   &       0.11  &       -73.6  &       1.9  &  7.7\\             
 76 & 05 35 25.10  &      -05 10 10.2   &           2.62   &       0.20  &       -63.8  &       2.2  &  2.9\\             
 77 & 05 35 23.86  &      -05 10 10.2  &            2.02   &       0.32   &       -48.1  &       4.4  &  1.8\\             
 78 & 05 35 27.58  &      -05 09 51.7  &            0.45  &       0.11  &       -59.2  &       6.8  &  7.6\\             
 79 & 05 35 25.10  &      -05 09 51.7  &            0.74  &       0.24  &       -42.6  &       9.6  &  2.4\\             
 80 & 05 35 23.86  &      -05 09 51.7   &           1.62   &       0.39  &       -41.9  &       6.8   &  1.5\\            
 81 & 05 35 30.06 &      -05 09 33.1    &          2.32    &      0.55   &      75.3    &      6.6   &  1.2\\            
 82 & 05 35 27.58 &      -05 09 33.1   &           0.65   &      0.09  &      -31.3   &      4.0   &  5.7\\            
 83 & 05 35 26.34  &      -05 09 33.1   &           1.20   &       0.14  &       -8.7  &       3.3  &  3.4\\             
 84 & 05 35 25.10  &      -05 09 33.1  &            1.57   &       0.23  &       -18.2  &       4.1    & 1.9 \\            
 85 & 05 35 23.86  &      -05 09 33.1  &            2.63   &       0.33  &       0.5  &       3.5   &  1.4\\            
 86 & 05 35 27.58  &      -05 09 14.6   &           1.12    &       0.14   &       -30.7  &       3.6   & 3.7 \\            
 87 & 05 35 26.34  &      -05 09 14.6   &           0.96  &       0.16  &       -0.4  &      4.8   &  2.9\\            
 88 & 05 35 25.10  &      -05 09 14.6   &           1.23   &       0.22  &       -15.5  &       5.0   & 2.0 \\            
 89 & 05 35 23.86  &      -05 09 14.6   &           2.09   &       0.34  &       10.4   &       4.6  &  1.4\\             
 90 & 05 35 22.61  &      -05 09 14.6   &           2.15   &       0.60  &       54.2   &       7.7  &  0.9\\             
 91 & 05 35 28.82  &      -05 08 56.1   &           2.88    &       0.61  &       -56.5   &      6.0  & 1.2 \\             
 92 & 05 35 27.58  &      -05 08 56.1   &           2.40   &       0.36  &       -51.6  &       4.1  &  1.8\\             
 93 & 05 35 26.34  &      -05 08 56.1   &           2.08    &       0.30  &       -16.3   &      4.2  &  1.7\\             
 94 & 05 35 25.10  &      -05 08 56.1   &           2.71   &       0.22  &       -7.8  &       2.4  &  2.1\\             
 95 & 05 35 23.86  &      -05 08 56.1    &          1.44   &       0.32  &       -8.5   &       6.0   &  1.8\\            
 96 & 05 35 22.61  &      -05 08 56.1    &          2.11   &       0.60  &       36.9   &       7.9  &  1.1\\             
 97 & 05 35 28.82  &      -05 08 37.5    &          2.33   &       0.56  &       -84.6  &       7.0  &  0.8\\             
 98 & 05 35 27.58  &      -05 08 37.5   &           1.54   &       0.38  &       -33.8   &      6.6  &  1.5\\             
 99 & 05 35 26.34  &      -05 08 37.5   &           2.39   &       0.30  &       -19.9   &      3.6  &  2.1\\             
 100 & 05 35 23.86  &      -05 08 37.5  &            0.94   &      0.25   &       2.9  &        7.5  &  2.6\\             
 101 & 05 35 22.61  &      -05 08 37.5  &            1.93   &        0.38  &       9.2  &        6.1  &  1.5\\             
 102 & 05 35 21.37  &      -05 08 37.5  &            3.70   &        0.73  &       31.9  &        5.6  &  1.1\\             
 103 & 05 35 27.58  &      -05 08 19.0  &            1.18  &        0.37  &       -23.9  &       8.6  &  1.6\\             
 104 & 05 35 26.34  &      -05 08 19.0  &            0.84  &        0.25  &       -70.1  &       8.2  &  2.3\\             
 105 & 05 35 25.10  &      -05 08 19.0  &            1.10  &        0.20  &       53.2   &       4.6  &  2.6\\             
 106 & 05 35 22.61  &      -05 08 19.0   &           1.29  &        0.31  &       24.0  &        6.7   &  1.8\\             
 107 & 05 35 21.37  &      -05 08 19.0   &           1.73  &        0.54  &       1.7  &        8.9   &  1.2\\            
 108 & 05 35 28.82  &      -05 08 00.4  &            2.09  &        0.56   &       -36.5   &      7.3  &  1.0\\             
 109 & 05 35 27.58  &      -05 08 00.4  &            3.25  &        0.44  &       -67.6  &       3.9   &  1.4\\             
 110 & 05 35 26.34  &      -05 08 00.4  &            1.91   &       0.36  &       -68.2  &       5.3  &  1.6\\             
 111 & 05 35 23.86  &      -05 08 00.4  &            1.30  &        0.17  &       -20.1    &      3.8  &  3.0\\             
 112 & 05 35 22.61  &      -05 08 00.4  &            1.97   &       0.26  &       -22.3  &       3.7  &  2.1\\             
 113 & 05 35 21.37  &      -05 08 00.4  &            1.92   &       0.51  &       -18.0  &       7.2  &  1.1\\             
 114 & 05 35 27.58  &      -05 07 41.9 &            4.90   &       0.59  &       -67.5     &       3.5  & 1.1 \\             
 115 & 05 35 26.34  &      -05 07 41.9  &            2.44  &        0.48  &       -64.6  &       5.4   &  1.2\\            
 116 & 05 35 23.86  &      -05 07 41.9  &            0.73  &       0.16  &       -45.0 &       6.7   &  3.2\\            
 117 & 05 35 22.61  &      -05 07 41.9  &            1.50  &        0.21  &       -67.8  &       4.0   &  2.1\\            
 118 & 05 35 21.37  &      -05 07 41.9  &            1.52  &        0.33  &       -1.7  &       6.2  &  1.2\\             
 119 & 05 35 27.58  &      -05 07 23.4  &            7.43  &        0.59  &       -51.3  &       2.3  &  1.0\\             
 120 & 05 35 26.34  &      -05 07 23.4  &            2.14  &        0.37  &       -56.3  &       4.8  &  1.2\\             
 121 & 05 35 25.10  &      -05 07 23.4  &            1.68    &        0.29  &       -41.9  &       4.6  & 1.5 \\             
 122 & 05 35 23.86  &      -05 07 23.4  &            0.73  &       0.17  &       -46.3  &       6.0  &  2.7\\             
 123 & 05 35 22.61  &      -05 07 23.4  &            1.27  &        0.21  &       -52.5  &       4.8  &  2.3\\             
 124 & 05 35 21.37  &      -05 07 23.4  &            1.09  &        0.31  &       -31.2  &       7.8  &  1.4\\             
 125 & 05 35 26.34  &      -05 07 04.8  &            5.58  &        0.52   &       -47.3   &       2.5  & 0.8\\             
 126 & 05 35 25.10  &      -05 07 04.8  &            3.25  &        0.36   &       -53.9  &       3.0  &  1.3\\             
 127 & 05 35 23.86  &      -05 07 04.8  &            2.22  &        0.21  &       -66.9  &       2.7  &  2.3\\             
 128 & 05 35 22.61  &      -05 07 04.8  &            1.62  &        0.24  &       -18.4  &       4.2  &  2.1\\             
 129 & 05 35 26.34  &      -05 06 46.3  &            6.14  &        0.56   &       -83.1  &       2.7  &  0.6\\             
 130 & 05 35 25.10  &      -05 06 46.3  &            5.99  &        0.55  &       -81.0  &       2.6     & 0.7 \\            
 131 & 05 35 23.86  &      -05 06 46.3  &            5.36  &        0.37  &       -71.3  &       1.9   &  1.2\\            
 132 & 05 35 22.61  &      -05 06 46.3  &            2.02  &        0.28  &       -70.4  &       3.8   &  1.8\\            
 133 & 05 35 21.37  &      -05 06 46.3  &            3.06  &        0.50  &       -61.0  &       4.9  &  1.1\\             
 134 & 05 35 22.61  &      -05 06 27.7  &            3.61  &        0.34  &       -80.8  &       2.5  &  0.8\\             
 135 & 05 35 21.37  &      -05 06 27.7  &            2.77  &        0.48  &       -81.7  &       4.5  &  0.5\\             
\enddata
\tablenotetext{(a)}{All the data shown here have a polarization level and uncertainty such that, $p/ \sigma_{P} > 3.$}.
\tablenotetext{(b)}{P.A. of E-vector in degrees east from north.}
\end{deluxetable}
  
\clearpage

\begin{deluxetable}{cccccccccc} 
\tablewidth{0pt}
\tabletypesize{\scriptsize}
\rotate
\tablecaption{Mean polarization percentages and position angles along OMC-3 and OMC-2. \label{MEANDISP}}
\tablehead{
  \colhead{Region$^{(a)}$}                  & \colhead{Vectors ID}              &
  \colhead{Number of}               & \colhead{$<p>$}                   & 
  \colhead{s(p)}                    & \colhead{$<\theta>$}              & 
  \colhead{s($\theta$)}             & \colhead{$<F>$}                   & 
  \colhead{s($F$)}                  & \colhead{$<$P.A.$_{\rm fila}>$}   \\
  \colhead{(see Figure \ref{polmap})}    & \colhead{(see Table \ref{DATA})}    & 
  \colhead{vectors}                 & \colhead{($\%$)}                  & 
  \colhead{($\%$)}                  & \colhead{$(^{\circ}$)}            & 
  \colhead{($^{\circ}$)}            & \colhead{$(Jy/18.6^{\arcsec}$ beam)}                   & 
  \colhead{$(Jy/18.6^{\arcsec}$ beam)}                  & \colhead{$(^{\circ}$)}            }
\startdata
OMC2/OMC3      & ...              &251 & 2.8  &    1.6  &   -34.6  &   35.6    &   1.6    &  1.0   &... \\
OMC3           & ...              &114 &3.5   &   1.7   &  -44.6   &   27.1    &   1.3    &  0.9   &... \\
OMC2	       &1:135             &135 &2.3   &   1.3   &  -21.9   &   39.9    &   1.7    &  1.2   &... \\
MMS1 to MMS6   & ...              &62  &3.4   &   1.8   &  -30.3   &   17.0    &   1.6    &  1.0   &-40 \\
MMS7           & ...              &17  & 2.9  &    1.3  &   -27.3  &    18.6   &   1.0    &  0.4   &-18 \\
MMS8, 9 $\&$ 10& ...              &37  & 3.8  &    1.8  &   -76.7  &    11.9   &   1.0    &  0.5   &23 or 113$^{(b)}$\\
FIR1 $\&$ 2    &94:135            &42  &2.5   &   1.6   &  -41.5   &   35.6    &   1.6    &  0.6   &-17 \\
FIR3, 4 $\&$ 5 &58:93             &36  & 2.0  &   1.0   &   -47.0  &    34.9   &   2.4    &  1.8   &-10 \\   
FIR6           &27:57             &31  & 1.9  &    1.1  &   26.2   &    44.5   &   1.6    &  0.7   &30 \\
South of FIR6  &1:26              &26  & 2.8  &   1.0   &   -5.7   &    24.1   &   1.2    &  0.4   &21 \\
\enddata
\tablenotetext{(a)}{Table of the OMC-3 data is available electronically.}
\tablenotetext{(b)}{See discussion in section \ref{OMC-3}}.
\end{deluxetable}

\clearpage

\begin{deluxetable}{cccccccc} 
\tablewidth{0pt}
\tabletypesize{\scriptsize}
\tablecaption{Results for the Power Indices, the Dispersions, the Turbulent-to-Mean Magnetic Field Strength Ratios, the Line Widths, 
and the Mean Field Strengths. \label{subregions}}
\tablehead{
\colhead{Region} &  \colhead{Number of} &
\colhead{$\gamma^{(a)}$} & \colhead{s($\theta$)} & 
\colhead{$b$} & \colhead{$<B_{t}^{2}>^{1/2}/B_{0}$} &
\colhead{$\sigma_{(v)}$} & \colhead{$B_{0}$} \\
\colhead{} & \colhead{pixels} &
\colhead{} & \colhead{$(^{\circ}$)} & 
\colhead{$(^{\circ}$)} & \colhead{}& 
\colhead{(km.s$^{-1}$)} & \colhead{(mG)} 
}\startdata
OMC-2/OMC-3    & 251 & ...              &35.6&21.2$\pm$0.8&...          &...         &...                      \\
OMC-3          & 116 & -0.40 $\pm$ 0.01 &27.1&12.1$\pm$1.0&...          &...         &...                      \\
OMC-2          & 135 & -0.66 $\pm$ 0.02 &39.9&28.2$\pm$0.8&...          &...         &...                      \\
MMS1 to MMS7   & 79  & -0.38 $\pm$ 0.01 &17.3&13.4$\pm$0.9&0.17$\pm$0.02&0.46$^{(b)}$&0.19                     \\
MMS8 to MMS10  & 37  & -0.90 $\pm$ 0.04 &11.9&11.1$\pm$1.2&0.14$\pm$0.02&...         &0.50$\times \sigma_{(v)}$\\
FIR1 and FIR2  & 42  & -1.08 $\pm$ 0.05 &35.6&22.9$\pm$0.8&0.30$\pm$0.02&...         &0.23$\times \sigma_{(v)}$\\
FIR3 to FIR5   & 36  & -0.47 $\pm$ 0.04 &34.9&26.0$\pm$0.8&0.34$\pm$0.02&0.65$^{(c)}$&0.13	               \\
FIR6 and south & 57  & -0.80 $\pm$ 0.05 &39.7&33.8$\pm$0.8&0.46$\pm$0.02&...         &0.15$\times \sigma_{(v)}$\\
\enddata
\tablenotetext{(a)}{Power indices obtained from plots similar to those shown in Figure \ref{pvsi}.} 
\tablenotetext{(b)}{Valid in OMC-3 MMS6, see Table 1 of \citet{hou00}.} 
\tablenotetext{(c)}{Valid in OMC-2 FIR4, see Table 1 of \citet{hou00}.} 
\end{deluxetable}

\clearpage

\begin{deluxetable}{lcccccc} 
\tablewidth{0pt}
\tabletypesize{\scriptsize}
\tablecaption{Jet/Outflow vs. Submm Polarization Orientations \label{JETS}}
\tablehead{
  \colhead{Flow$^{(a)}$}     & \colhead{R.A.(J2000)}         &
  \colhead{Dec.(J2000)}      & \colhead{P.A.$_{\rm flows}$}   &
  \colhead{$<{\rm P.A.}_{\rm submm}>^{(b)}$} & 
  \colhead{$\Delta $P.A.} \\
  \colhead{Designation}                & \colhead{($^{\rm h}$ $^{\rm mn}$ $^{\rm s}$)}& 
  \colhead{($^{\circ}$ $\arcmin$ $\arcsec$)}   & \colhead{$(^{\circ})$}           &      &
  \colhead{$(^{\circ})$}        & \colhead{$(^{\circ})$}           }
\startdata
mms23              &5 35 18.5&-5 00 28&87    &-33& 60&\\
5/mms5             &5 35 22.4&-5 01 16&90/88 &-35& 56&\\
4                  &5 35 23.4&-5 01 31&-4    &-28& 24&\\
mms7               &5 35 26.5&-5 03 50&70    &-17& 87&\\
mms8               &5 35 26.5&-5 05 18&65    &-63& 52&\\
mms9$^{(\rm west)}$&5 35 26.0&-5 05 47&80    & 89& 9 &\\
mms9$^{(\rm east)}$&5 35 26.0&-5 05 47&80    &-65& 35&\\
mms10              &5 35 32.2&-5 05 47&34    &-66& 80&\\
14                 &5 35 28.1&-5 07 20&45    &-59& 76&\\
fir1bc             &5 35 23.4&-5 07 48&-2    &-53& 51&\\
fir2               &5 35 24.5&-5 08 30&-15   &-13& 2&\\
18                 &5 35 27.5&-5 09 17&56    &-22& 78&\\
19                 &5 35 26.7&-5 09 24&-83   &-28& 55&\\
17/fir3            &5 35 27.5&-5 09 37&31    &-57& 72&\\
21$^{(\rm north)}$ &5 35 27.2&-5 11 11&-1   &-61& 60&\\
21$^{(\rm south)}$ &5 35 27.2&-5 11 11&-1   & -3& 2&\\
23$^{(\rm north)}$ &5 35 22.8&-5 11 50&18    &-58& 76&\\
23$^{(\rm south)}$ &5 35 22.8&-5 11 50&18    & 28& 10 &\\
24                 &5 35 23.3&-5 12 03&61    &-66& 53&\\
25                 &5 35 21.4&-5 13 14&-20   & 32& 52&\\
\enddata
\tablenotetext{(a)}{
\citet{wil03} used abbreviations fir and mss followed by numbers to identify CO outflows.
\citet{sta02} used single numbers to identify H2 jets. We use their designations here.} 
\tablenotetext{(b)}{Vectors lying at a distance less than 18.6" from the projected Jet/Outflow axis are used. 
No vector is used more than one time.}
\end{deluxetable}

\end{document}